\documentclass[12pt]{article}

\usepackage{jheppub}
\makeatletter
\def\@fpheader{\relax}
\makeatother

\usepackage{caption}
\usepackage{subcaption}
\usepackage{amsmath,amssymb,amsfonts,graphicx,slashed,color,amsthm,mathtools,upgreek}
\usepackage{hyperref}
\usepackage{float}
\usepackage[font=small ,labelfont=bf ]{caption}
\usepackage{csquotes}
\usepackage{tikz}
\usepackage{footnotebackref}

\newtheorem{definition}{\bf DEFINITION}

\begin{document} 

\title{${}$\\\vspace{.3in}
Dynamics of genetic code evolution: The emergence of universality \footnote{The author list is alphabetical.} }

\author[a]{John-Antonio Argyriadis}
\author[b]{\!\!, Yang-Hui He}
\author[c]{\!\!, Vishnu Jejjala}
\author[d]{\!\!, Djordje Minic}

\emailAdd{john-antonio.argyriadis@jesus.ox.ac.uk}
\emailAdd{hey@maths.ox.ac.uk}
\emailAdd{vishnu@neo.phys.wits.ac.za}
\emailAdd{dminic@vt.edu}

\affiliation[a]{Jesus College, University of Oxford, OX1 3DW, UK;\\
Rudolf Peierls Centre for Theoretical Physics, Clarendon Laboratory, Parks Rd, University of Oxford, OX1 3PU, UK}
\affiliation[b]{Department of Mathematics, City, University of London, EC1V 0HB, UK;\\
Merton College, University of Oxford, OX1 4JD, UK;\\
School of Physics, NanKai University, Tianjin, 300071, P.R.~China}
\affiliation[c]{Mandelstam Institute for Theoretical Physics, School of Physics, NITheP, and CoE-MaSS, University of the Witwatersrand, Johannesburg, WITS 2050, South Africa;\\
David Rittenhouse Laboratory, University of Pennsylvania, Philadelphia, PA 19104, USA}
\affiliation[d]{Department of Physics and Center for Soft Matter and Biological Physics, Virginia Tech, Blacksburg, VA 24061, USA }

%\emailAdd{john-antonio.argyriadis@jesus.ox.ac.uk}
%\emailAdd{hey@maths.ox.ac.uk}
%\emailAdd{vishnu@neo.phys.wits.ac.za}
%\emailAdd{dminic@vt.edu}

\abstract{
We study the dynamics of genetic code evolution.
The model of~\citet{nigel} and~\citet{vestigian} uses the mechanism of horizontal gene transfer to demonstrate convergence of the genetic code to a near universal solution.
We reproduce and analyze the algorithm as a dynamical system.
All the parameters used in the model are varied to assess their impact on convergence and optimality score.%achieving universality.
We show that by allowing specific parameters to vary with time, the solution exhibits attractor dynamics. %converges much faster to a universal solution.
Finally, we study automorphisms of the genetic code arising due to this model.
We use this to examine the scaling of the solutions in order to re-examine universality and find that there is a direct link to mutation rate.
}

\date{}

\renewcommand{\baselinestretch}{1.15}
\parskip=.4\baselineskip

\newcommand\be{\begin{equation}}
\newcommand\ee{\end{equation}}
\newcommand\bea{\begin{eqnarray}}
\newcommand\eea{\end{eqnarray}}
\newcommand\eref[1]{(\ref{#1})}
\newcommand\nn{\nonumber}

\newcommand\comment[1]{}
\newcommand\vj[1]{\noindent{\color{red}{\textbf{[#1 -V.]}}}}
\newcommand\herebedragons{\begin{center}{\LARGE \color{blue}{\textbf{--- HIC SUNT DRACONES ---}}}\end{center}}

\maketitle

\setcounter{footnote}{1}

\section{Introduction}\label{intro}
The genetic code arose through evolution.
We think of it as being universal, optimal, and highly redundant~\cite{Sella2006}.
The mechanics of the evolution of life on Earth means that all organisms share the same genetic code in which each codon produces the same amino acid.
We call this the \textbf{standard genetic code}.
(While there are deviations from the standard genetic code~\cite{exception}, these involve only a handful of codons being coded differently and will not be the subject of our investigations here.)
Because of selection pressures, the standard genetic code has self optimized to minimize errors in translation and  transcription~\cite{1inamillion,Freeland2003,Haig1991}.
From theoretical considerations, \citet{woese} showed that the standard genetic code is related to a property called the \textit{polar requirement}, which has subsequently been corroborated by experiment~\cite{Mathew2008} and shown to be highly optimal when considering one type of error: point mutations~\cite{butler}.
It can be considered as part of an abstract chemical property of the genetic code~\cite{sella}.
Following~\citet{nigel} and~\citet{vestigian}, in this paper, we consider the code's optimality in terms of these features.

\citet{sella} attempted to model the evolution of the genetic code.
This was done through considering the coevolution between the genetic code and the encoding of a protein within a closed model system.
This allows for complex dynamics between mutations of messages % \footnote{Messages are genes that code for a protein. A gene is a DNA sequence that creates a single protein.}
and selection on proteins in order to minimize the lethal effects of these mutations.
%Such dynamics allows the evolutionary barrier to be overcome in a protein dominated world~\cite{nigel}.
%\vj{I have no idea what the previous sentence means. just cut it-JAA}
This minimizes the errors through mutations and allows protein networks to develop to promote a higher likelihood of survival.
We seek to understand whether the algorithm based on this coevolutionary model can be phrased as a purely physical problem of dynamical evolution.
To address this we must first discuss the algorithm.

The algorithm described by~\citet{nigel} models the evolution of the genetic code through \textbf{horizontal gene transfer (HGT).}
This allows organisms to exchange genetic information via DNA through transferring the segments of a genome to each other within the same generation through various mechanisms.
This is used in concert with the Code Message Coevolution Model dynamics described by~\citet{sella} to obtain an iterative discrete time algorithm.
Implementation of the algorithm demonstrates convergence of the genetic code to a highly optimised and near universal solution.
This solution is an attractor. %highly optimised and universal configuration.
Horizontal gene transfer is crucial to achieving this solution.
This result provides a testable model for understanding the standard genetic code.

The dynamics from the coevolution model along with the additional communication incorporated by the iterative discrete time algorithm governs the evolution of genetic code states (genetic code configurations) of which there is a finite number.
This allows us to treat the algorithm as a discrete dynamical system, as it is the time dependent dynamics of coevolving matrices which represent the genetic code state.
We will consider this algorithm as if it were a system out of equilibrium for which there is the emergence of an attractor solution in the space of genetic code mappings~\cite{strogatz2000nonlinear}.

The behavior we establish exemplifies the notion of \textit{universality}.
In statistical physics, systems with a large number of degrees of freedom exhibit universality in a scaling limit.
Historically, this idea originated in the theory of phase transitions and was made mathematical precise through the renormalization group.
(See, for example,~\cite{Wilson:1974mb,Wilson:1993dy,Goldenfeld:1992qy}.)
Starting from specific initial conditions, by integrating out degrees of freedom (\textit{e.g.}, through coarse-graining), we flow to a fixed point.
At the fixed point, the theory is scale invariant.\footnote{
This is a simplification.
At a fixed point of the renormalization group, the theory enjoys a larger symmetry, \textit{conformal invariance}, which includes scale or dilatation invariance.}
Very different physical systems can flow to the same fixed point in that correlation functions of local operators behave in an identical manner.
Such theories are said to belong to the same universality class.
Input parameters dictate how the system converges to universality.
It is therefore natural to examine how variations on these parameters influence the universality of the solution.
We will use approximate scale invariance as a tool to assess how close we are to universality and to diagnose features of the universal solution.

In this article, we investigate a mechanism for the origin of the genetic code that leads to universal behavior at late times.
We initially describe the model with considerations of universality as in \citet{nigel} by defining it in such a way that all entities within the algorithm converge to a single solution. 
We then consider universality in a more formal manner through statistical mechanics.

Let us begin by stating the main results.
The first requirement for the convergence of the genetic code is a trivial observation that we make rigorous:
there must be more codons than amino acids. 
The second requirement is that we must demand horizontal gene transfer to optimize the setup.
This corroborates the claims of~\citet{nigel}.
We also discover that universality in terms of scaling in the solution depends on the rate of mutations. % and the time necessary to achieve mutational equilibrium. 
 It is largely independent of mistranslations of the genetic code.

The organization of this paper is as follows.
In Section~\ref{sec:two}, we recast the biological algorithm into a computational algorithm to which we can apply the principles of dynamical systems.
We reproduce the results from~\citet{nigel} and show that the initial conditions do not influence the algorithm's ability to flow to a near universal solution.
Our new results are in Section~\ref{sec:three}.
In particular, in Section~\ref{subsec:31}, we correct minor errors in the literature.
In Sections~\ref{subsec:32}--\ref{subsec:34}, we vary all the parameters in the model in order to examine their influence on the attractor mechanism. %algorithm's ability to yield a universal solution.
In Section~\ref{subsec:36}, we discuss automorphisms and scaling in the genetic code.
We then illustrate the mechanism for universality in terms of scaling with an example.
In Section~\ref{subsec:38}, we re-examine universality and express this behavior in terms of the homogeneity of the fitness function.
We find that the universal solution is characterized by the rate of mutations and is largely independent of the mistranslation rate.
In Section~\ref{sec:four}, we conclude with a summary and directions for future work.
Finally, the Appendices collect the results of various experiments less central to the argument than those discussed in the main text.

\section{Modeling framework}\label{sec:two}
\numberwithin{equation}{section}

We begin with a precise rephrasing of the problem addressed by~\citet{nigel} into one of computation.
Emphasis will be on representing the aspects of the mathematical problem while minimizing the amount of biology introduced.

\subsection{Basic definitions}\label{subsec:21}

We model the set of codons making up DNA geometrically using a Hamming metric~\cite{baylis2017error}.

\begin{definition}\label{codons}
Let $i_b \in i $ be a set of elements (\textbf{bases}) forming an alphabet of length $|i|$.
We define a \textbf{codon} as a sequence of $n$ bases such that $c \in \mathcal{C}:=\{i_{b,1},\dots , i_{b',n} \}$.
The number of possible codons is $|\mathcal{C}| = |i|^n$.
\end{definition}
For codons in the standard genetic code, we have $|i|=4$ (A,C,G,T) and $n=3$ meaning that $|\mathcal{C}|=64$.
We define the set of codons, $\mathcal{C}$, lexicographically.
Note that there is an associated symmetry with a Hamming metric~\cite{automorph}.

We can next define the structure of the genetic code:
\begin{definition}\label{code}
Denoting the set of animo acids $a \in \mathcal{A}$ such that we have $|\mathcal{A}|$ amino acids, the mapping from codon space to amino acid space, $G : \mathcal{C} \rightarrow  \mathcal{A}$ is the \textbf{genetic code}.
We represent the map $G$ as matrix $\Delta_{c,a}$ with dimensions $ |\mathcal{C}| \times |\mathcal{A}|$ such that:
\begin{equation}
\Delta_{c,a}= \left\{
\begin{array}{lcl}
1 && \text{if}\ G(c) = a ~,\\
0 && \text{otherwise} ~.
\end{array} \right.
\label{eq:code}
\end{equation}
\end{definition}

We refer to $\Delta_{c,a}$ as the \textbf{delta matrix}.
This matrix defined in \eqref{eq:code} has one entry per row (as each codon can only map to a single amino acid) and no empty columns (we assume that every amino acid has been mapped to).
The map $G$ is surjective but not injective.
In Nature, we encounter $20$ amino acids in the genetic code.\footnote{
For the purposes of calculation, we treat the stop codons as mapping to a dummy amino acid, so in our language $|\mathcal{A}|=21$ in the standard genetic code.}
The codon TTT maps to the amino acid phenylalanine, for example.
This matrix can therefore be considered as a non-square, row-stochastic, binary matrix with no empty columns.
Notice that when summing over the columns of a row-stochastic matrix, we get $1$.
These properties place constraints on the information flow for optimization~\cite{bialek2012biophysics}.
Note that due to these constraints, we also must have $ |\mathcal{C}|\geq |\mathcal{A}|$.
%As the matrix is generically not invertible, this corroborates the central dogma of biology. 
Using the inclusion/exclusion principle~\cite{Allenby:2010:CIC:1942905}, the number of possible configurations of the delta matrix $\Delta_{c,a}$ is
\begin{equation}
\#_{\text{config}}= \sum_{j=0}^{|\mathcal{A}|}  (-1)^j  \binom{|\mathcal{A}|}{j}(|\mathcal{A}|-j)^{|\mathcal{C}|} \ .
\label{eq:config}
\end{equation}

During the map $G$, errors occur with a given probability:
\begin{equation}
\text{prob}(c \rightarrow a )= \sum_{c'} L_{c,c'}\Delta_{c',a}\ , \qquad 
L_{c,c'}(\ell)=\left\{
\begin{array}{lcl}
\frac{\ell}{(n(|i|-1))} && \text{dist}(c,c')=1 ~, \\
1-\ell && \text{dist}(c,c')=0 ~, \\
0 && \text{otherwise} ~. \
\end{array}\right.
\label{eq:2.3}
\end{equation}
The parameter $\ell\in[0,1] \subset \mathbb{R}$.
The distance is defined using the Hamming metric such that:
\begin{equation}
\text{dist}(c,c') := \#(c_j\neq c'_j)_{j=1,\dots, n} \ .
\label{eq:2.4}
\end{equation}
The distance is the number of bases $i_b$ that differ between two codons $c$ and $c'$.
Here, $\ell$ represents a parameter for the probability of error, and $L_{c,c'}(\ell)$ is a bistochastic matrix, \textit{viz.}, a symmetric, non-negative matrix whose rows and columns sum to $1$.
The matrix $L_{c,c'}(\ell)$ is used in order to only consider nearest neighbors in codon space.
The number of codons with $\text{dist}(c,c')=1$ is given by  $n(|i|-1)$ as there are $n$ positions which can have $|i|-1$ different values.
We can encode this information in a \textit{Hamming graph} in which the $\mathcal{C} = |i|^n$ possible codons correspond to vertices and an edge joins vertices whose corresponding codons that differ by a single letter --- \textit{i.e.}, those codons at Hamming distance $1$.

In this algorithm the genetic code $G$ rearranges itself in order to minimize the likelihood that probabilistic nature of the map causes a differing amino acid $a$ to appear when mapping from codon space~\cite{Freeland2003,Haig1991}.
In this model we consider two forms of errors, both of which only occur on nearest neighbors ($\text{dist}(c,c')=1$).
They are the following:
\begin{enumerate}
\item \textbf{Mistranslation}: When a single base $i_b$ is read incorrectly.
We will denote this $T_{c,c'}$ and take $\ell \rightarrow \nu$, where $\nu$ is the rate of mistranslation.
\item \textbf{Point mutations}: A single base $i_b$ changes before being read.
We will denote this $M_{c,c'}$ and $\ell \rightarrow \mu$ where $\mu$ is the rate of mutations.
There are various kinds of point mutations.
\end{enumerate}
For simplicity, we will neglect excisions or insertions of bases.

\subsection{Fitness}\label{subsec:22}
Information is translated from genome to proteome.
For our purposes, these are sequences of codons and amino acids, respectively.
In particular,
\begin{definition}\label{genome}
A sequence $S_G = \{c_1, \ldots, c_M\}$ of length $M$ is called a \textbf{genome}, where each codon $c_x \in \mathcal{C}$ has a position $x$ in the sequence $\{1,\dots, M\}$.
A target amino acid $s(x)$ is the mapping under the genetic code $G$ of the codon at position $x$ to the amino acid $s(x) \in \mathcal{A}$.
The image under the genetic code map $G$ of the genome sequence $S_G$, gives a sequence $S_P =\{s_1, \ldots, s_M\}$ of target amino acids called the \textbf{proteome}, which is a subsequence of a protein. 
\end{definition}
We denote the target amino acid $s(x)$ as $s$ in order to abbreviate notation.
The definition we quote above is a slight simplification of~\cite{nigel,vestigian,sella} as in this algorithm, we assume $s \equiv a \in \mathcal{A}$.
As the amino acids at each position in the sequence are indistinguishable \cite{nigel}, we can store details of the proteome and genome within the following objects:
\begin{definition}\label{codeusage}
A vector $L_s$ specifies the frequency of the target amino acid $s$ in a proteome sequence of length $M$.
The codon usage matrix $U_{c,s}$ specifies the frequency of a codon $c$ within a target amino acid $s$.
\end{definition}
Crucially, we encode all necessary information about the genome and proteome within these two objects without having to go through the respective sequences analytically.
The two matrices are both column stochastic, \textit{i.e.},
\begin{equation}
\sum_{s}^{|\mathcal{A}|} L_{s}= 1 ~, \qquad \sum_{c}^{|\mathcal{C}|} U_{c,s} = 1_s ~.
\label{eq2.5}
\end{equation}
\newpage
The notion of distance in amino acid information space is structurally ambiguous (not well defined like a Hamming metric).
Due to this we can define the topological distance between amino acids by the following $a_d \in [0,1]$, which can be randomly generated.
The notion of distance is normalized.
Using this we can define a \textbf{fitness matrix} as:
\begin{equation}
W_{a,s}= \Phi^{|a_d - s_d|} ~.
\label{eq2.6}
\end{equation}
As in~\citet{sella}, $\Phi$ is a parameter used to consider how an abstract physicochemical distance between amino acids scales into the fitness. %abstract chemical distance between amino acids.
This makes the fitness matrix \eqref{eq2.6} some measure of how \enquote{useful} each arbitrary amino acid $a$ is instead of the target amino acid $s$.
Since $0 < \Phi \leq 1$, this is a positive symmetric matrix.
By considering the probability of mistranslations and the entire genome we can describe an overall fitness score~\cite{nigel}:
\begin {equation}
f=\prod_c \prod_s \{ \sum_{c'} \sum_{a}T_{c,c'} \Delta_{c',a}W_{a , s} \}^{ L_{s} U_{c,s}} \ .
\label{eq2.7}
\end{equation}
This product is taken component wise.

To measure how well a delta matrix performs, we define the \textbf{optimality score} $O$ as:
\begin{equation}
O= \sum_c \sum_{c'} (N_{c, c'}  \{ \sum_a \sum_b \Delta_{c,a}S_{a,b}\Delta_{c',b}^T\}) ~,
\label{eq2.8}
\end{equation}
which measures the average amino acid similarity between neighboring codons.
We define amino acid similarity as $S_{a,b}=\sum_s |W_{a , s}-W_{s,b}|$.
In~\eqref{eq2.8}, $N_{c, c'}$ is $1$ if two codons are nearest neighbors ($\text{dist}(c,c')=1$) and zero otherwise~\cite{nigel,vestigian}.
Note that it is a tautology that two isomorphic genetic codes give the same optimality score. 
We return to this point in Section~\ref{subsec:36} below.

\subsection{The algorithm}\label{subsec:23}
Based on these mathematical preliminaries, we consider the following algorithm~\cite{nigel}.
\begin{enumerate}
\item \textbf{Construction:}
We can construct a set of $N$ objects each with their own genetic code $G$ and therefore delta matrix $\Delta_{c,a}$ and their own codon usage matrix $U_{c,s}$.

\item \textbf{Mixing:} We randomly select one object as the acceptor $A$ and a random subset $K$ of $N$ as the donors ($k \in K\subset N$) and run them through the iteration:
\begin{equation}\
( 1 - H)  U_{c,s}^A +  \frac{H}{K}\sum_{k \in K} U_{c,s}^{(k)} \to   U_{c,s}^{A} ~.
\label{eq2.7}
\end{equation}
$H$ represents the fraction of the genetic code due to horizontal gene transfer ($H\in[0,1]$).

\item \textbf{Fitness maximization:} We attempt an elementary code change to the delta matrix $\Delta_{c,a}$.
We do this by assigning one codon to a new amino acid.
This is done by reallocating a unit entry in $\Delta_{c,a}$ to a different position within that row of the matrix.
We accept the new code if and only if it preserves or increases the fitness score $f$, which has been calculated using the new $U_{c,s}^{A}$.
Otherwise, we keep the original delta matrix $\Delta_{c,a}$, if there are no new possibilities.

\item \textbf{Mutational equilibrium:} We can derive a new codon usage matrix $U_{c,s}^{A}$ from the new delta matrix $\Delta_{c,a}$ uniquely at mutational selection equilibrium.
We first derive a fitness matrix with respect to codons $F_{c,s}=\sum_{a}\Delta_{c,a}W_{a,s}$.
Using the Perron--Frobenius theorem, we calculate the column stochastic eigenvector corresponding to the largest eigenvalues, for the following matrix ($Q^s$):
\begin{equation}
Q_{c,c'}^s =\sum_{c''}M_{c,c''}\delta_{c'',c'}F_{c'',s} ~,
\label{eq:2.8}
\end{equation}
where $\delta_{c'',c'}$ is a Kronecker delta so that we consider the $s^{th}$ column of the matrix $F_{c'',s}$ as a diagonal matrix.
The index $s$ here is fixed and not a free index.
Each column stochastic eigenvector of $Q_{c,c'}^s$ corresponds to the $s^{th}$ column of $U_{c,s}^{A}$.
We normalize the eigenvector so that it is column stochastic) by setting the sum of elements to unity.

\item \textbf{Repetition:} We repeat steps $2$ through $4$ for $t$ time steps.
\end{enumerate}

\paragraph{Experimental setup:}
In this model there are $12$ parameters to generate and define.
These are tabulated as follows.
\begin{itemize}
\item \textbf{Space structure:}
$|i|$ and $n$ for the codon space,
$|\mathcal{A}|$ and $a_d$ for amino acid space,
and $L_s$ target amino acid frequency;
\item \textbf{Innovation pool structure:}
$N$ number of objects,
$K$ number of donors per iteration, and
$H$ fraction of genome that is similar due to horizontal gene transfer;
 \item \textbf{Noise and fitness parameters:}
 $\nu$, $\mu$, and $\Phi$;
 \item \textbf{Number of time steps:} $t$.
 \end{itemize}

\begin{figure}[H]
\centering
\begin{tabular}{c c}
\includegraphics[width=0.5\linewidth]{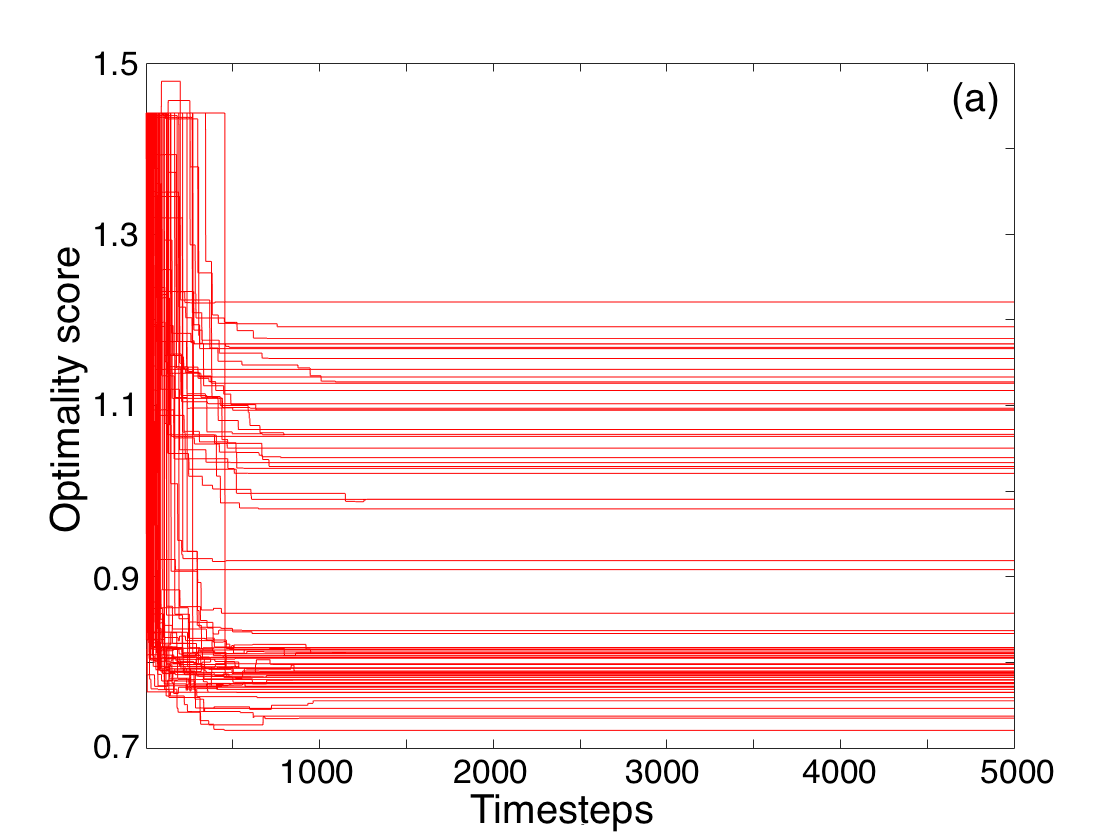}
\includegraphics[width=0.5\linewidth]{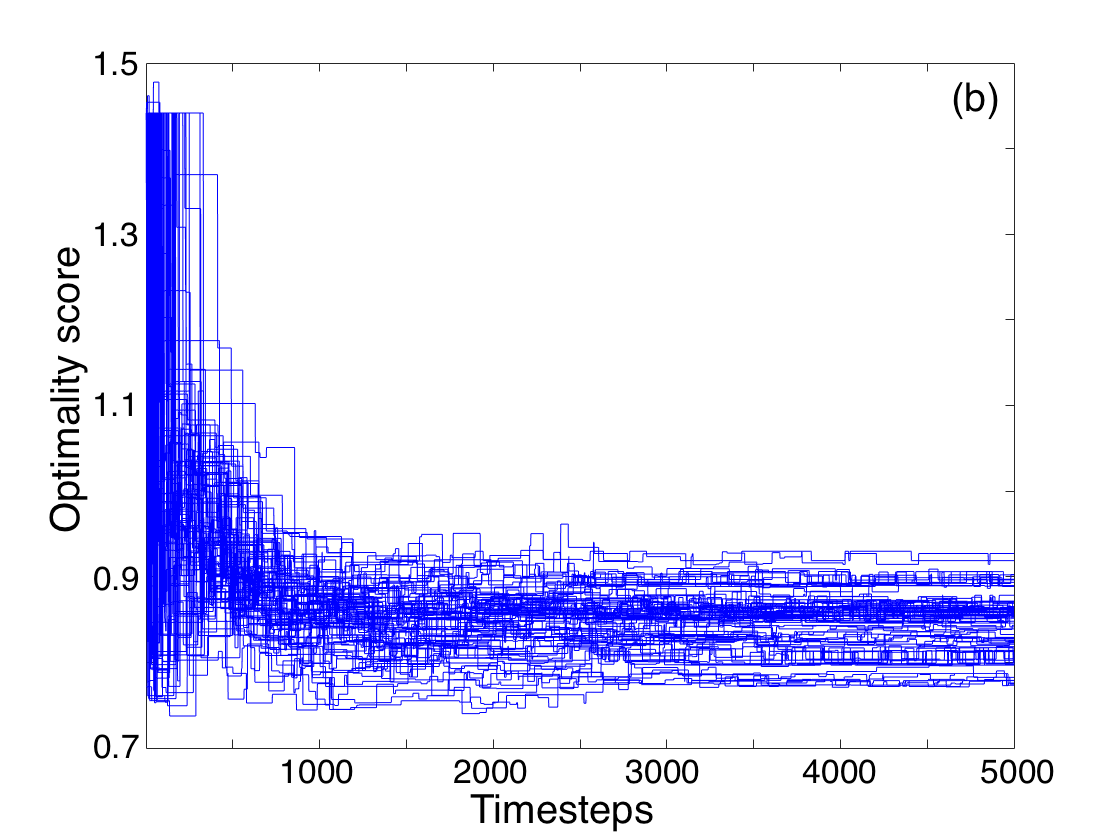}
\end{tabular}
\caption{Evolution of optimality score for (a) $H=0$ (\textcolor{red}{red}) and (b) $H=0.4$ (\textcolor{blue}{blue}).
Both graphs were produced using the following parameters: $|i|=4$, $n=3$ giving $|\mathcal{C}|=64$, $|\mathcal{A}|=20$, $N=80$, $K=1$, $\nu=0.01$, $\mu=10^{-4}$, and $\Phi=0.99$.
$L_s$ and $a_d$ are the same for both graphs.}
\label{fig:one}
\end{figure}

We generate  $L_s$ and $a_d$ randomly.
The parameters $|i|$, $n$, $|\mathcal{A}|$, $N$, and $K$ are positive integers while $H$, $\nu$, $\mu$, and $\Phi$ take any value in the interval $[0,1]$.
We reproduce the results in~\citet{nigel} using their parameters as quoted in Figure~\ref{fig:one}.
The initial delta matrices $\Delta_{c,a}$ are identical at the start, as done in \citet{nigel}.%with the assumption that all entities originate from the last universal common ancestor (LUCA), a theoretical construct which suggests that all life on earth originated from a single organism (or single population of organisms).
The results in Figure~\ref{fig:one} show that when modeling without horizontal gene transfer (Figure \ref{fig:one} (a), red), the delta matrices $\Delta_{c,a}$ optimize themselves, but do not converge to a universal solution.
This is shown by the optimality scores, $O$, ranging from $0.7$ to $1.25$ and not changing after $1500$ time steps.
When including horizontal gene transfer (Figure \ref{fig:one} (b), blue) we get a set of optimality scores, $O$, that optimize on average more than without horizontal gene transfer (red, $H=0$).
The results display the attractor mechanism. %will eventually converge.
This is because the optimality score falls in a smaller range (between $0.75$ and $1$) and fluctuations continuing at the the $t=5000$ time step.

The time taken to produce these results was very large as we use $|i|=4$, $n=3$, and $|\mathcal{A}|=20$ giving us $64\times20$ matrices.
To perform a more careful analysis, we consider a toy model by reducing the matrix dimensions to $27\times9$.
This corresponds to the parameters $|i|=3$, $n=3$ (so $|\mathcal{C}|=27$) and $|\mathcal{A}|$=9.
We also set up the algorithm so that each entity has its own unique delta matrix $\Delta_{c,a}$ such that they all start with different initial optimality scores in order to see if the scores will still converge.
These results are in Appendix~\ref{app:a}.
They show some convergence after $5000$ time steps.
This suggests that the set of $\Delta_{c,a}$ can be arbitrary in order for optimised attractor mechanism to emerge. %an universal and optimal solution to emerge.
This also points to the existence of an attractor mechanism at work for $H\ne 0$. %a universal solution as long as there is some \enquote{attractor mechanism} at work for $H\ne 0$.

When performing the analysis all initial delta matrices $\Delta_{c,a}$ will be identical.
% The results for $|i|$, $n$ and $|\mathcal{A}|$ are in Appendix~\ref{app:b}; for $N$, $K$ in Appendix~\ref{app:c}, and finally, for $\nu$, $\mu$ and $\Phi$ in Appendix~\ref{app:d}. Note that variations in $H$ are given in Section~\ref{subsec:33}.
We will vary a single parameter from the set $\{|i|, n, |\mathcal{A}|, N, K, H, \nu, \mu, \Phi \}$ while keeping the others  fixed.
We generate $a_d$ and $L_s$ randomly for all runs. 
We will take ten runs for each parameter and take the average.
The standard deviation will be used to analyze the spread (to measure the rate of convergence).
We use the standard deviation as it allows us to measure the spread of optimality scores in an intuitive manner, similar to the mean code distance used in \citet{nigel} as we highlight further in Section~\ref{subsec:36}.
We will take the average of the standard deviation over the ten runs.
These results are discussed in Sections~\ref{subsec:32}--\ref{subsec:34} below, where we consider universality to be represented by the ability for all entities within the algorithm to converge to a single solution.
The remainder of Section~\ref{sec:three} is devoted to re-defining and re-examining universality using statistical mechanics and the renormalisation group.

\section{Results and analysis}\label{sec:three}

\subsection{The model of Sella and Ardell}\label{subsec:31}

In an insightful paper, \citet{sella} develop a Code Message Coevolution Model that describes the impact of message mutation on the fitness of the genetic code.
The authors observe that at mutational equilibrium, there is a balance between mutations in messages and selection on proteins.
This model has been summarized in~\citet{ardell}. 
The process involves calculating the column stochastic eigenvector corresponding to the largest eigenvalues as described in step~4 (mutational equilibrium) of the algorithm described in Section~\ref{subsec:23}.
We note for completeness and reproducibility two minor errata.
Example~A from~\citet{sella} consists of a model with the following setup.
There is a ring of five codons mapping to a ring of five amino acids with $\Delta_{c,a}$ being the identity matrix.
Note that we will denote the eigenvector as $U_{c,s}$, however, $s$ is fixed and is not a free index (as in our step~4). 
To reproduce the results, we take $\Phi = 0.8^5$ and $\mu=0.01$.\footnote{
We are grateful to D.~Ardell for communications on this point.}
The resulting eigenvalues and eigenvectors are given in Table~\ref{tab:one}.
Note that at machine precision the eigenvectors sum to one ($\sum_c U_{c,s}=1$) as required.

\begin{table}[H]
\begin{center}
\par\medskip
\textbf{Eigenvalues and column stochastic eigenvectors for a set of $\Phi$ and $\mu$ }
\begin {tabular}{c c c c}
&
Scaling for an abstract  
&
Rate of mutations $\mu$
&
Largest eigenvalue $\lambda^s$
\\
&
physicochemical between amino acids $\Phi$
\\ \hline
\textbf{1}
&
0.8 
&\
0.1 
& 
0.9549
\\ 
\textbf{2}
&
$0.8^5$
&
0.01
&
0.9808
\end{tabular}
\begin{tabular}{c c}
&
Corresponding eigenvector ${U_{c,s}}^T$
\\[1ex] \hline 
\textbf{1}
&
$\begin{bmatrix}
0.2635 & 0.2134 &  0.1549 &  0.1549 &  0.2134
\end{bmatrix}$
\\[1ex]
\textbf{2}
&
$\begin{bmatrix}
0.9058 & 0.0461 & 0.0011 & 0.0011 &0.0461\\
\end{bmatrix}$
\end{tabular}
\caption{Row \textbf{1} gives the resulting eigenvalues and eigenvectors for the parameters stated in the paper.
Row \textbf{2} provides the eigenvalues and eigenvectors from the paper using the corrected parameters.
Note these eigenvectors are for a given value of $s$. }
\label{tab:one}
\end{center}

\end{table}

\subsection{Varying parameters for space structure}\label{subsec:32}

Recall that $|i|$ counts the number of nucleotides.
These are the letters that comprise a DNA sequence.
Taking a codon to consist of an $n$ nucleotide sequence, the number of possible codons is then $|\mathcal{C}| = |i|^n$.
These codons describe $|\mathcal{A}|$ amino acids.
With four bases as in the standard genetic code, there are $64$ three base sequences corresponding to possible codons.
These codons correspond to $20$ amino acids, so we have a many to one map.
In this subsection, we report on experiments involving varying parameters corresponding to the spatial structure of the map.

\newcounter{blah}
\setcounter{blah}{1}

\begin{figure}[h!!!]
\centering
\begin{tabular}{c c}
\includegraphics[width=0.47\linewidth]{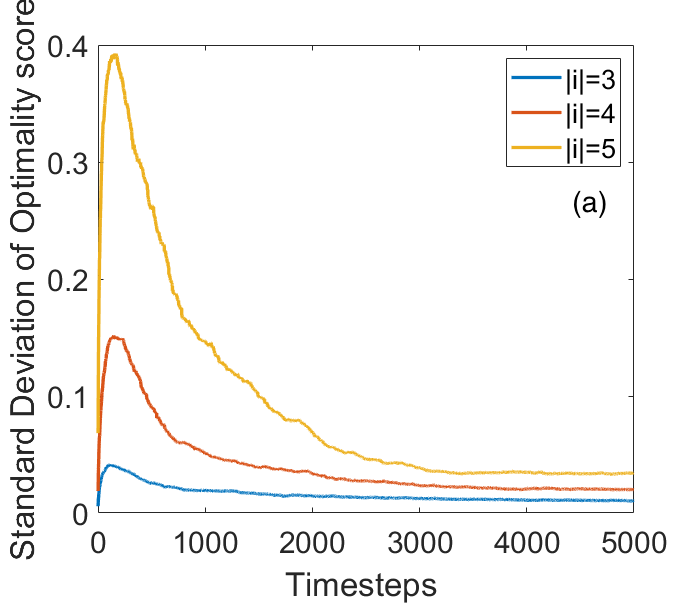}
&
\includegraphics[width=0.5\linewidth]{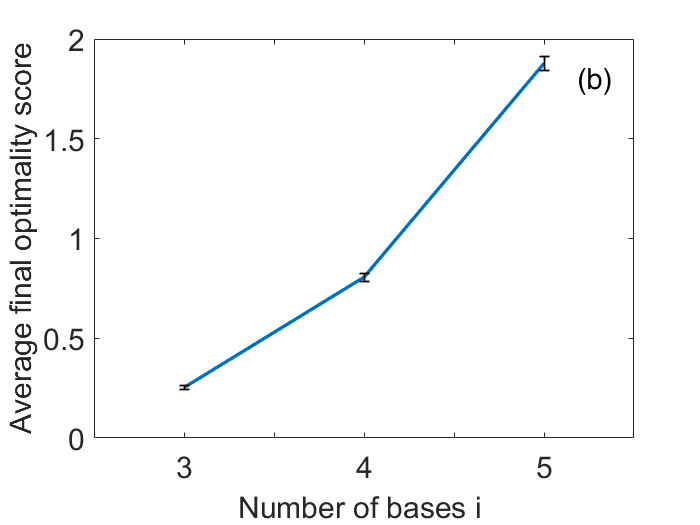}
\\
&
\end{tabular}
\caption{Figure \ref{fig:base} (a) shows the average time evolution of the standard deviation of the optimality score for a given $|i|$ over ten runs.
Figure \ref{fig:base} (b) shows the average final optimality score for a given $|i|$ over ten runs.
The initial parameters are the same for all runs: $n=3$, $|\mathcal{A}|=9$, $N=80$, $K=1$, $H=0.4$, $\nu=0.01$, $\mu=10^{-4}$, and $ \Phi=0.99$.
The error bars show the average one standard deviation spread of final optimality scores over the ten runs (to measure the rate of convergence).}
\label{fig:base}
\end{figure}

\paragraph{Experiment~\theblah: Varying the number of nucleotides}${}$\\
We test the fitness optimization for the cases $|i|= 3 ,4 ,5$. 
We do not take $|i|=2$ as this gives $|\mathcal{C}|=8<|\mathcal{A}|=9$, breaking one of the constraints on the delta matrix.
We do not take $|i| \geq 6$ either as this produces a matrix that is at minimum $216\times 9$, which takes significant processing time to iterate.
The results are given in Figure~\ref{fig:base}.
For all values of $|i|$, Figure~\ref{fig:base} (a) displays an optimised solution with an attractor mechanism converging to a near universal solution.%implies that the optimality score will eventually converge to a universal solution. 
When varying $|i|$ we find the value of the optimality score, $O$, increases proportionally as shown in Figure~\ref{fig:base} (b).
This makes sense as increasing $|i|$ increases $|\mathcal{C}|$ meaning we sum over more elements to get the optimality score $O$.
The rate of convergence decreases as indicated by the error bars increasing proportionally with $|i|$ in Figure~\ref{fig:base} (b).
This is as expected as larger matrices should take longer to find the universal solution.

% \vj{Fix $x$-axis on Figure~\ref{fig:base} (right).}

\stepcounter{blah}

\begin{figure}[H]
\centering
\begin{tabular}{c c}
\includegraphics[width=0.47\linewidth]{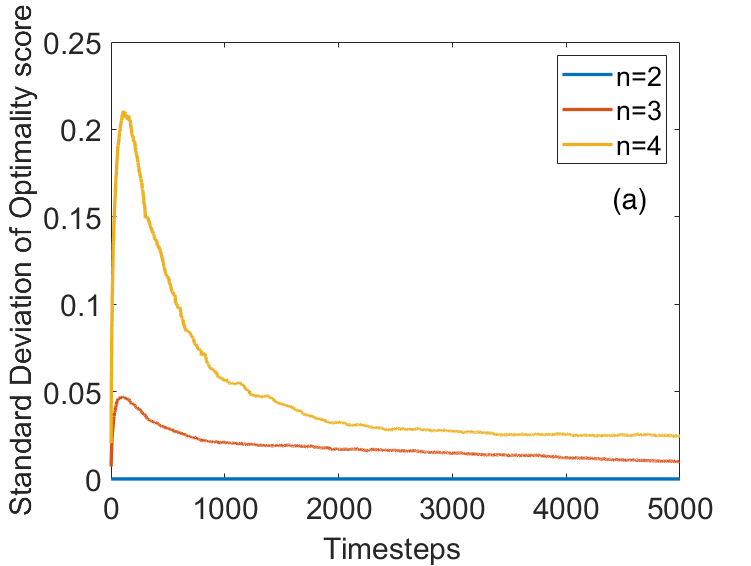}
&
\includegraphics[width=0.5\linewidth]{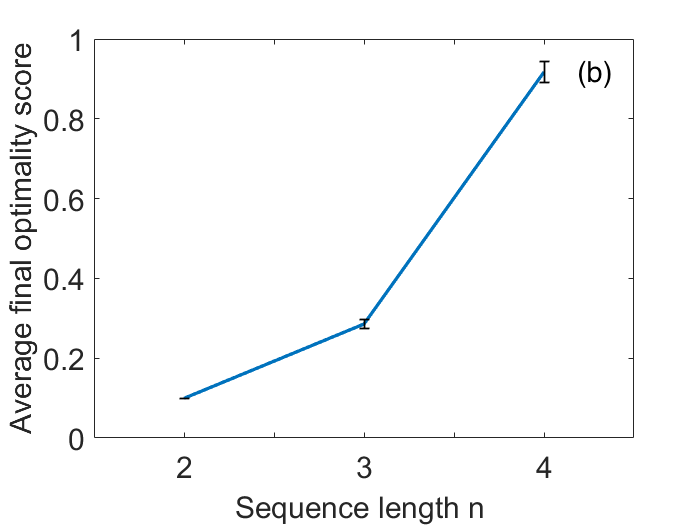}
%\includegraphics[width=0.5\linewidth]{}
%\\
%\includegraphics[width=0.5\linewidth]{bifurcationdiaginn.png}
%&
\end{tabular}
\caption{Figure \ref{fig:sequence} (a) shows the average time evolution of the standard deviation of the optimality score for a given $n$ over ten runs.
Figure \ref{fig:sequence} (b) shows the average final optimality score for a given $n$ over ten runs.
The initial parameters are the same for all runs: $|i|=3$, $|\mathcal{A}|=9$, $N=80$, $K=1$, $H=0.4$, $\nu=0.01$, $\mu=10^{-4}$, and $ \Phi=0.99$.
The error bars show the average one standard deviation spread of final optimality scores over the ten runs (to measure the rate of convergence).}
\label{fig:sequence}
\end{figure}

\paragraph{Experiment~\theblah: Varying the length of a codon}${}$\\
For the length of a codon $n$, we take $n=2, 3, 4$.
We do not take $n=1$ or $n\geq5$ for the same reasons as when varying $|i|$.
The results are given in Figure~\ref{fig:sequence}.
They display the same pattern as when varying $|i|$, because we are increasing $|\mathcal{C}|$ again.
When $n=2$, we get $|\mathcal{C}|=|\mathcal{A}|=9$ which means $\Delta_{c,a}$ forms a permutation matrix.
This matrix cannot be changed in step~3 of the algorithm discussed in Section~\ref{subsec:23} (fitness maximization) as we cannot reassign a single codon $c$ to a new amino acid $a$ without being left with an empty column.
The delta matrix, $\Delta_{c,a}$, cannot therefore evolve, giving a single flat line for $n=2$ as seen in Figure~\ref{fig:sequence} (a).
This implies that we require $|\mathcal{C}| > |\mathcal{A}|$ for the algorithm to work.
We take note of the smallness of the standard deviations in Figure~\ref{fig:sequence} (b).

% \vj{Fix $x$-axis on Figure~\ref{fig:sequence} (right).}

\stepcounter{blah}

\begin{figure}[h!!!]
\centering
\begin{tabular}{c c}
\includegraphics[width=0.5\linewidth]{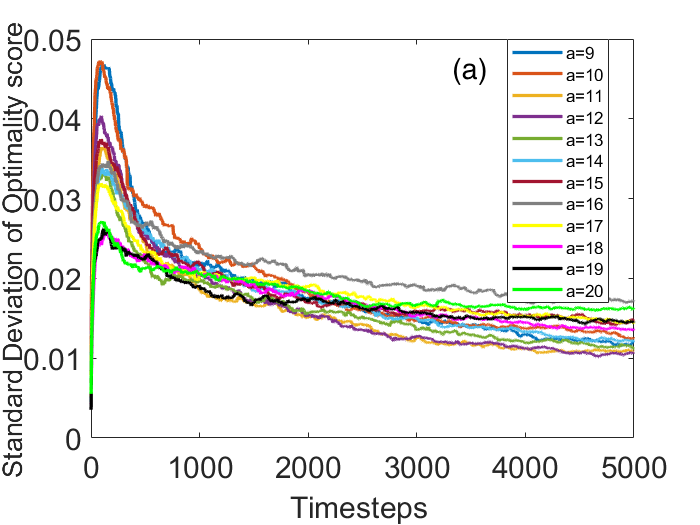}
&
\includegraphics[width=0.5\linewidth]{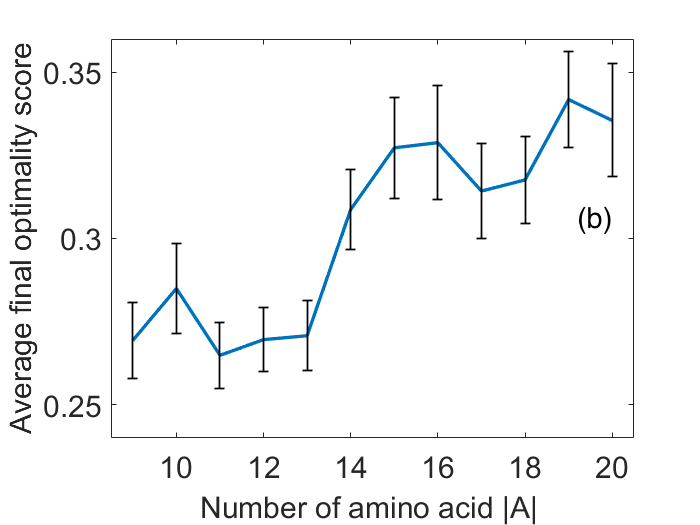}
\end{tabular}
\\
\caption{Figure~\ref{fig:acid} (a) shows the average time evolution of the standard deviation of the optimality score for a given $|\mathcal{A}|$ over ten runs. 
Figure~\ref{fig:acid} (b) shows $|\mathcal{A}|$ (number of amino acids) against the average final optimality score, averaged over ten runs.
The initial parameters are the same for all runs: $|i|=3$, $n=3$, $N=80$, $K=1$, $H=0.4$, $\nu=0.01$, $\mu=10^{-4}$, and $ \Phi=0.99$.
The error bars in show the average one standard deviation spread of final optimality scores over the ten runs (to measure the rate of convergence).}
\label{fig:acid}
\end{figure}

\paragraph{Experiment~\theblah: Varying the number of amino acids}${}$\\
The result of this experiment is that variations on $|\mathcal{A}|$ display convergence.
It is important to consider that the randomly generated values for topological amino acid distance $a_d$ and site frequency $L_s$ will also vary, as they are generated with consideration on $|\mathcal{A}|$.
The results for this are given in Figure~\ref{fig:acid}.
There appears to be an upwards trend in Figure \ref{fig:acid} (b). 
This result is intuitively expected as increasing the number of amino acids increases the number of terms we sum over; however, further investigation is needed to confirm this.
This should be done keeping randomly generated variables fixed where possible.

% \vj{Fix caption on Figure~\ref{fig:acid} (right).}

\stepcounter{blah}

% \vj{In this subsection, the captions state three runs and ten runs.
% I assume the latter is what is meant and have edited accordingly.}

\subsection{Varying parameters for the innovation pool structure}\label{subsec:33}
Recall that the algorithm from Section~\ref{subsec:23} begins by constructing $N$ objects each with a genetic code $G$.
At each time step, one of these objects receives a fragment of genome from $K$ donors selected from the set of objects.
The parameter $H$ computes the fraction of the recipient genome due to horizontal gene transfer.

\pagebreak
\paragraph{Experiment~\theblah: Varying $N$}${}$\\
The results for varying $N$ is shown in Figure~\ref{fig:entity} in Appendix~\ref{app:b}.
The number of entities $N$ is taken from $10$ to $100$ in steps of $10$.
The results when varying the number of entities show a linear relationship between number of entities $N$ and final average final optimality score as seen in Figure~\ref{fig:entity} (b).

% \vj{Can you do $N=90, 100$ as well?
% I would like to understand the blip at the end.}

\stepcounter{blah}

\paragraph{Experiment~\theblah: Varying $K$}${}$\\
When varying the number of donors $K$ from $1$ to $5$, we find that it does not affect the algorithm's dynamics as seen in Figure~\ref{fig:donor} in Appendix~\ref{app:b}.
This makes sense as we are always adding the same fraction $H$ to the acceptor codon usage $U_{c,s}^A$. 
We should note there does appear to be a slight upwards trend in average final optimality score as shown in Figure~\ref{fig:donor} (b). 
Further investigation should be undertaken with larger value of $K$ such that $K\approx N$.

\stepcounter{blah}

\paragraph{Experiment~\theblah: Varying $H$}${}$\\
We take $H$, the fraction of the genome similar due to horizontal gene transfer from $0$ to $1$ in increments of $0.1$.
The results are shown in Figure~\ref{fig:hgt} below.
Figure~\ref{fig:hgt} (a) that for $H=0$, there is no convergence as expected, while for $H=0.2$ the results begin to converge but at a very slow rate.
Looking at the Figure~\ref{fig:hgt} (b), it is clear that $0.4  \leq H \leq 0.7 $ gives the minimal optimality score and the smallest error bars.
This implies the final results have been substantially optimised and have converged to a greater extent via the attractor mechanism to a near universal solution.  %are both more optimal and are converging faster to the universal solution.
When $H\geq 0.8$ , the final scores, $O$, are less optimal and have converged less.%slower at converging.
This is probably  due to a change from \enquote{mixing} to \enquote{swapping} of codon usage matrices $U_{c,s}$, preventing optimal communication.
Results from Figure~\ref{fig:hgt} suggest that maximum mixing occurs around $H=0.6$.

%The results for varying $H$ the fraction of the genome similar due to HGT are given in appendix J where we take H from 0 to 1 in 0.1 steps. We can see clearly that H=0 (graph-J.1) shows no convergence of the results while H=0.1 (graphJ.2) shows the results beginning to converge but at a very slow rate. Looking at the distribution of the final optimality scores  given by figure 2, its clear that for H between 0.4 and 0.7 give both the minimal optimality score and the smallest range in standard deviations implying the final results is both more optimal and that the rate of convergence is much greater. For $H \geq 0.8$(graphs- J.9-J.11), the final results are probably less optimal and slower at converging due a change from 'mixing' to 'swapping' of codon usage matrices $U_{c,s}$ not allowing optimal communication. The maximum mixing occurs when $H=0.5$. 
%Further investigation of the results for a very high number of iterations for $0.1 \geq H \geq 0$ is needed to understand how the fixed point vary as the results imply some Bifurcation of the fixed points around H=0. We can interpret H as some measure on the 'attractive forces between entities. This  'attractive force' prevents any entities from getting stuck in any less optimal metastable states.

\begin{figure}[h!!!]
\centering
\begin{tabular}{c c}
\includegraphics[width=0.5\linewidth]{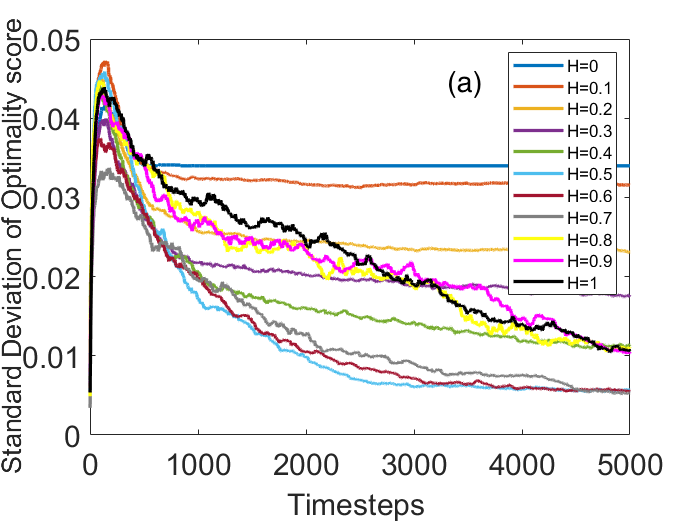}
&
\includegraphics[width=0.5\linewidth]{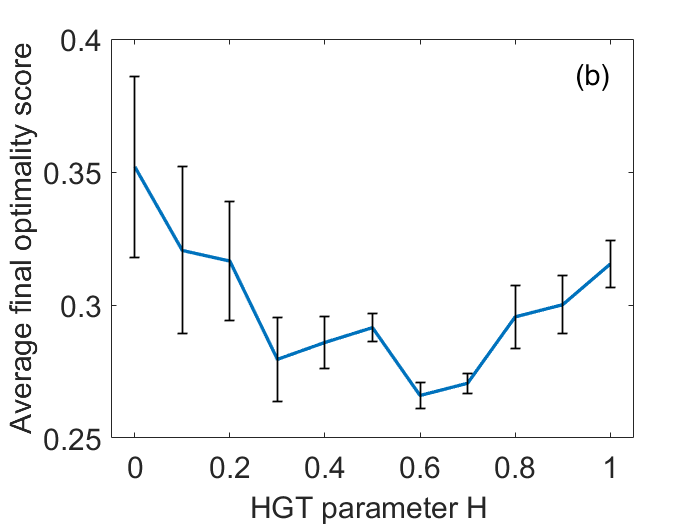}
\end{tabular}
\caption{The average final optimality score for a given $H$.
Figure~\ref{fig:hgt} (a) shows the average time evolution of the standard deviation of the optimality score for a given $H$ over ten runs.
Figure~\ref{fig:hgt} (b) shows the average final optimality score for a given $H$ over ten runs.
The initial parameters are the same for all runs: $|i|=3$, $n=3$, $|\mathcal{A}|=9$, $N=80$, $K=1$, $\nu=0.01$, $\mu=10^{-4}$, and $ \Phi=0.99$.
The error bars show the average one standard deviation spread of final optimality scores over ten runs (to measure the rate of convergence).}
\label{fig:hgt}

\end{figure}

\stepcounter{blah}

\paragraph{Experiment~\theblah: Time evolution of $H$}${}$\\
The parameter $H$ is best considered a variable that decreases with time \cite{Woese8742, nigel}. 
This is due to better translation of the model allowing evolution of a protein network with more specific interactions to occur \cite{nigel}. 
To model this, we define $H$ in the following manner:
\begin{equation}
H(t)=H_0e^{-kt} ~.
\label{eq:timeh}
\end{equation}

\begin{figure}[h!!!]
\centering
\begin{tabular}{c c}
\includegraphics[width=0.5\linewidth]{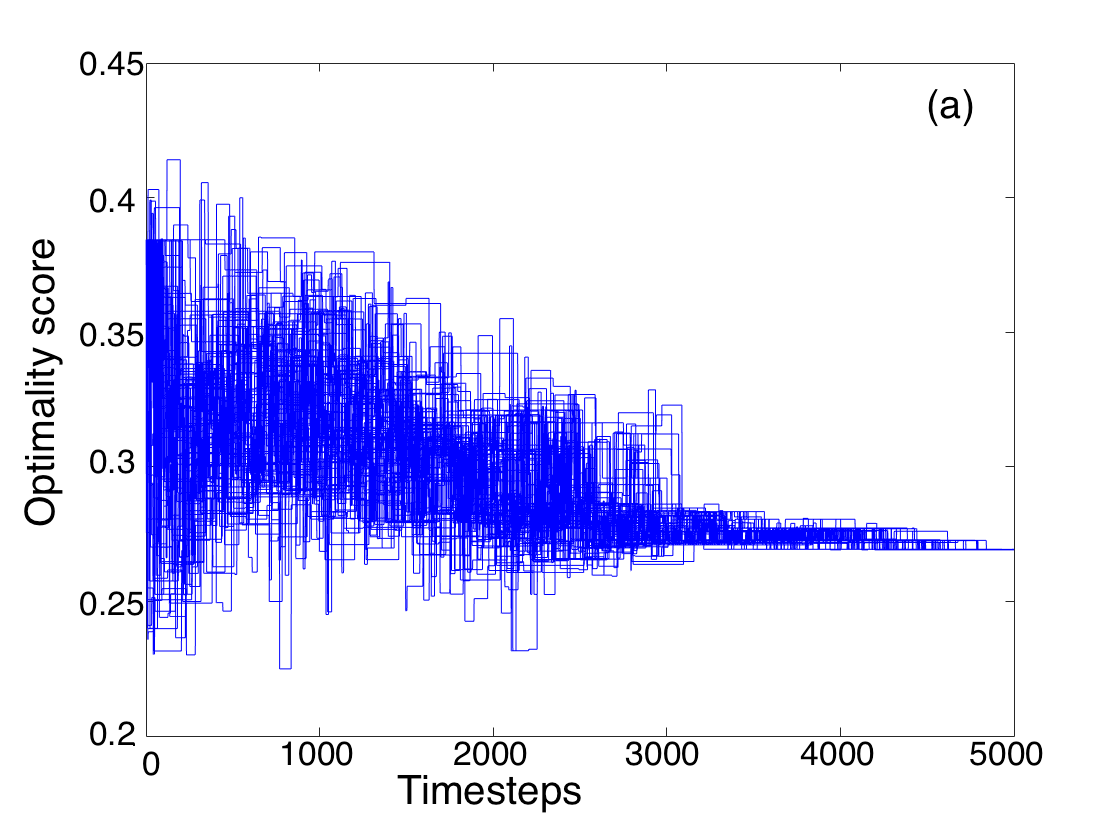}
&
\includegraphics[width=0.5\linewidth]{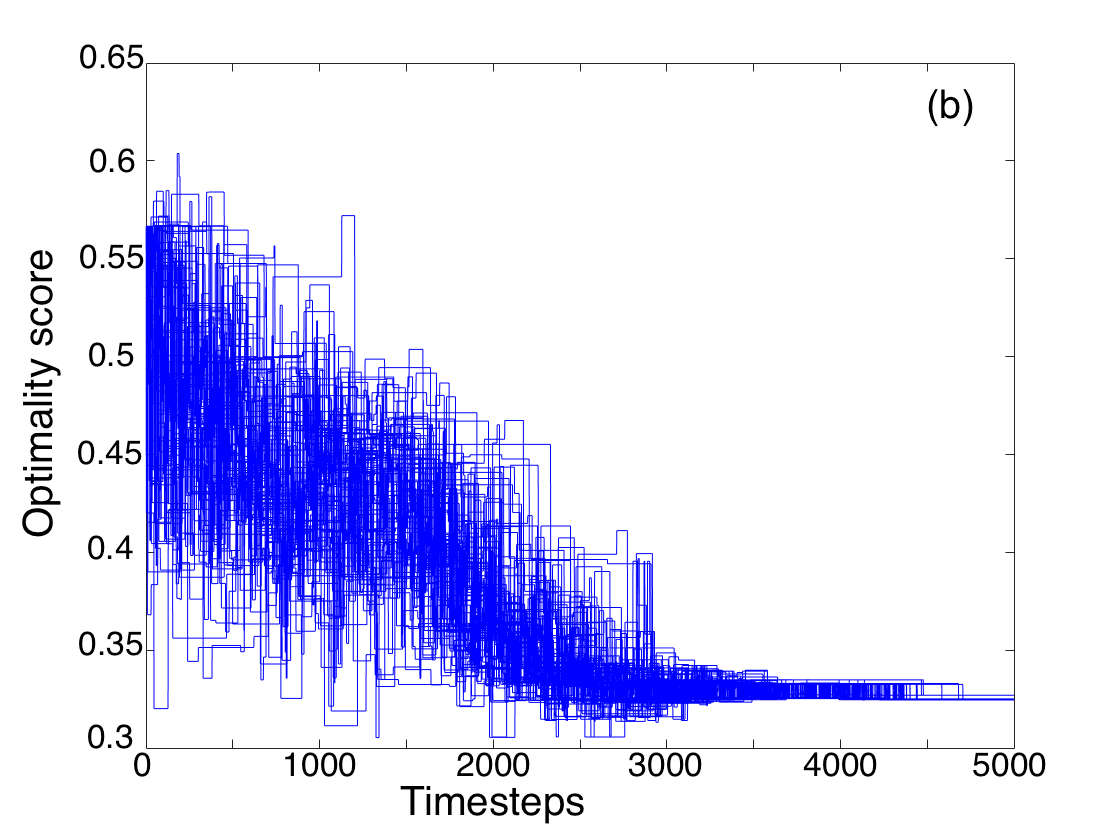}
\\
\includegraphics[width=0.5\linewidth]{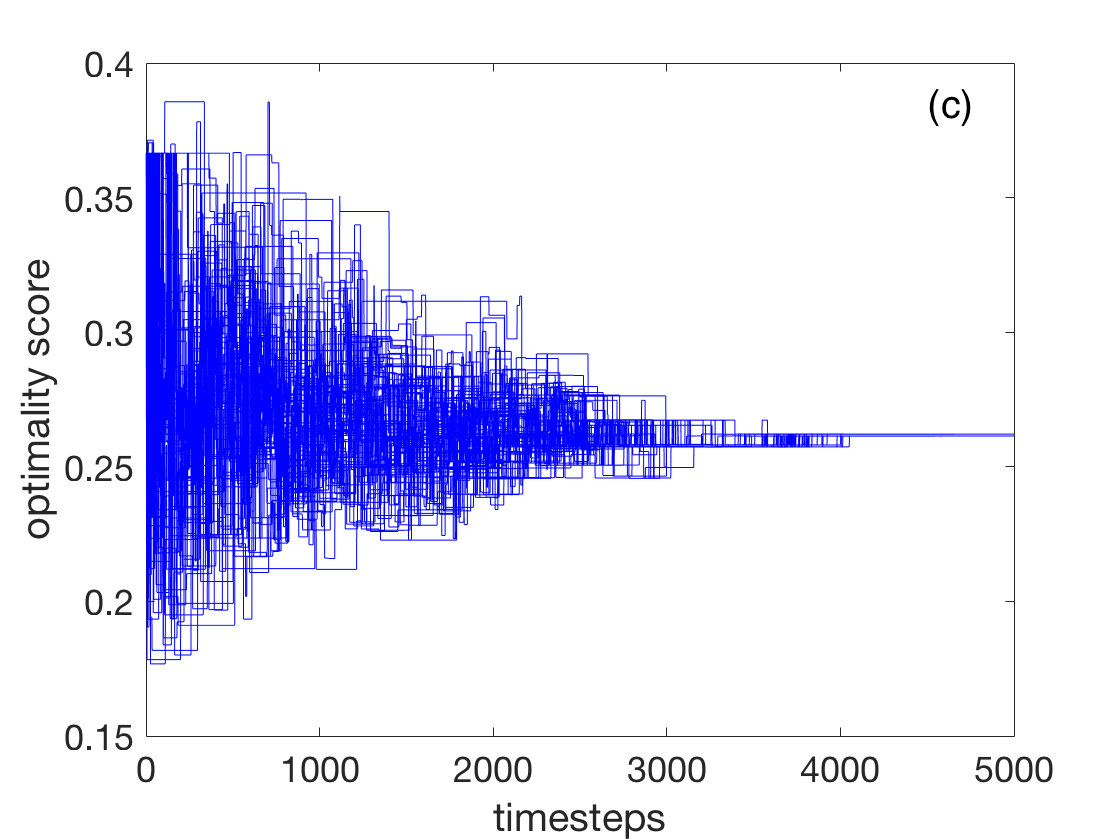}
&
 \includegraphics[width=0.5\linewidth]{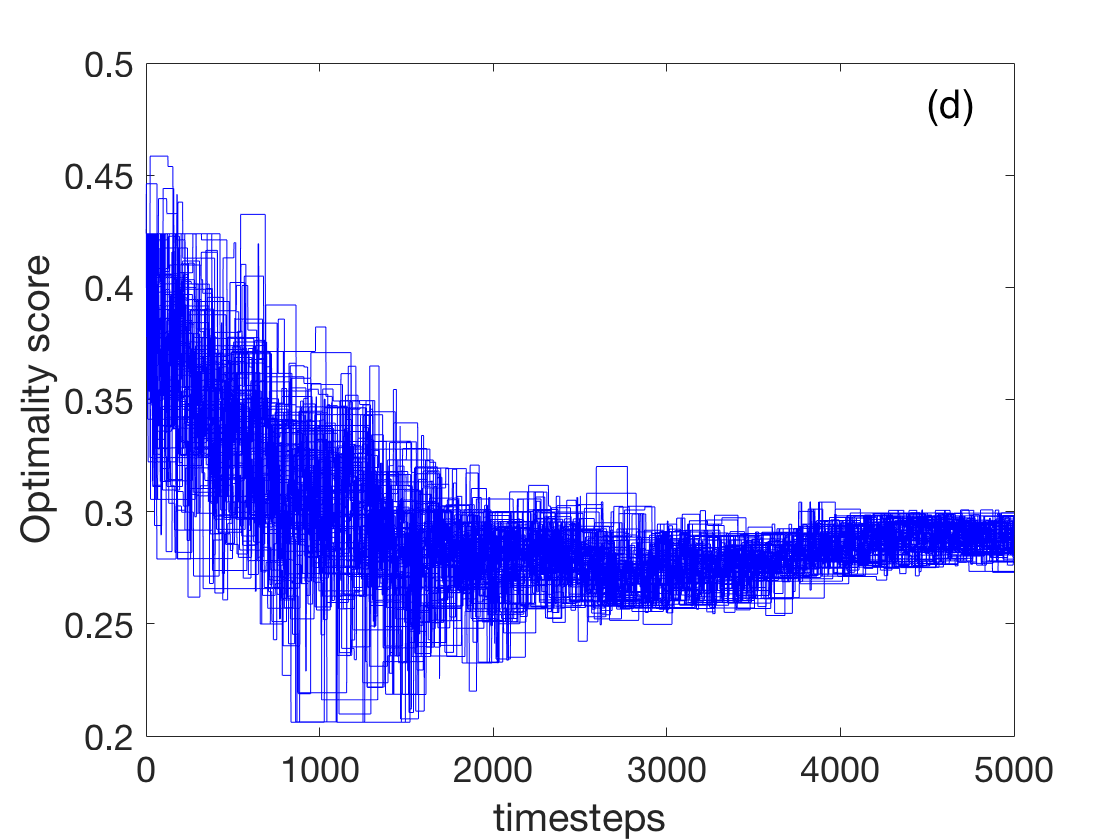}
\end{tabular}

\caption{Evolution of optimality score with a time evolving parameter $H$.
Four runs showing the evolution of optimality score for a time evolving $H$ (according to \eqref{eq:timeh}).
We use: $|i|=3$, $n=3$, $|A|=9$, $N=80$, $K=1$, $\nu=0.01$, $\mu=10^{-4}$, $ \Phi=0.99$, $k=10^{-4}$, and $H_0=1$.}
\label{fig:timeH}
\end{figure}

In this equation, $H_0$ is the initial fraction of horizontal gene transfer similar ($0\leq H_0\leq1$), and $k$ is a constant.
Initially, $H_0$ is set to $1$.
Setting $k=10^{-3}$ gives a number that is approximately zero at, say, $t=5000$.
The results in Section~\ref{sec:three} indicate that we expect convergence to occur after $5000$ iterations.
For this reason, we set $k=10^{-4}$ so that $H(t=5000)$ is relatively far from zero. 
The resulting dynamics is given in the Figure~\ref{fig:timeH} and Appendix \ref{app:evo}, where the degree of convergence is significantly improved in several runs, to all prior results.
The majority of the trials have fully converged and can be considered universal. %are essentially universal.
Note that the solutions converge to different values, due to $a_d$ being randomly generated.
We can see that only Figure \ref{fig:timeH} (d), does not completely converge.
The rate of convergence seem to to be fairly similar to the runs when using large constant $H$.
This makes sense as this is a stochastic process meaning the rate of convergence should vary between runs.
However, the average rate of convergence significantly improves in the cases where universality manifests within this time frame.
Note that $H(t)$ sits in the optimal range suggested in Section~\ref{subsec:33} for approximately the last $1500$ time steps.
\vspace{0.2cm}

\stepcounter{blah}

\pagebreak
\subsection{Varying error and fitness parameters}\label{subsec:34}
Recall that the parameters $\nu$, $\mu$, and $\Phi$ take values in the interval $[0,1]$.
The parameter $\nu$ measures the rate of mistranslation, when a single base is misread.
The parameter $\mu$ measures the rate of mutation, when a single base is changed.
The parameter $\Phi$ characterizes  how an abstract physicochemical distance between amino acids scales into the fitness. %the consequences of mistakes by providing a measure on the space of amino acids.

\paragraph{Experiment~\theblah: Varying $\nu$ and $\mu$}${}$\\
The plots for these variations are in Appendix~\ref{app:c}.
It can be seen that variations on $\nu$ have no effect on a given optimality score.
As we are trying to minimize the effects of errors from $\nu$ and $\mu$, there is less requirement to optimize as they decrease.
This can be seen in Figure~\ref{fig:error} (b) in Appendix~\ref{app:c}.
As these parameters decrease the optimality score increases.
Note we take $\nu$ and $\mu$ from $1$ to $10^{-4}$ on a log scale.
We do not try $\nu,\mu=0$ as this implies there is no need to optimize the code as no errors can occur.

\stepcounter{blah}

\paragraph{Experiment~\theblah: Varying $\Phi$}${}$\\
As described by \citet{vestigian}, $\Phi$ is a scale for the fitness for one amino acid substitution.
This implies that it should not affect the rate of convergence directly.
However, it will affect the score converged to.
To examine this we reduce the fitness score $f$ to a function of $\Phi $ and $\nu$ in order to consider their role in the algorithm given by Figure~\ref{fig:phi} (a) in Appendix~\ref{app:c}.
The rest of the values are randomly generated.
The variations of $\Phi$ are proportional to $f$ as expected.

\subsection{Re-defining universality within a genetic code model}\label{subsec:36}
The framework we have described so far shows that there is some degree of convergence via an attractor mechanism.%convergence to a single solution.
With horizontal gene transfer turned on ($H\ne 0$), we have an attractor.
While the details of the solution depend in part on the initial conditions assigned to parameters in the model, the model exhibits near universality at late times.
This is demonstrated by the converging behaviour of the optimality scores $O$ of all entities.%convergence of the optimality scores $O$ of all entities to a single solution.
We aim to refine the concept of universality.
To do this we must first understand genetic code configurations and the possible symmetries associated with them.
We will then analyze the fitness landscape of all genetic code configurations in order to to see if this function can be scaled homogeneously.
To re-examine universality further, we  mainly consider the model provided by \citet{sella} while also incorporating the fitness function provided by \citet{nigel}. 
% This does not explicitly address the algorithm and does not consider possible scaling with respect to time of the algorithm.}

Note that for this section we will also work with the fitness in the following form: 
\begin {equation} \label{eq3.2}
\log f=\sum_c \sum_s L_{s} U_{c,s} \log ( \sum_{c'} \sum_{a}T_{c,c'} \Delta_{c',a}W_{a , s} ) ~.
\end{equation}
The logarithm simplifies algebraic manipulations.
%This ensures everything is somewhat linear.

Using Definition~\ref{code}, we represent each genetic code configuration using a delta matrix $\Delta_{c,a}$.
We can also calculate total number of configurations using~\eqref{eq:config}.
This framework allows us to consider the genetic code mapping as a surjective mapping from a Hamming graph (of codons) to a random graph (of distance between amino acids in an abstract topological information space).
These graphs have automorphisms due to labeling which we will highlight clearly in an upcoming example.
The automorphisms imply that certain genetic code configurations (and therefore delta matrices $\Delta_{c,a}$) are isomorphic to each other, meaning that they represent the same genetic code map $G$ even though they have different delta matrices $\Delta_{c,a}$.
Considering the random graph is randomly generated, we \textit{a priori} assume that no automorphisms exist within the amino acid graph.
Note this is only true for $|\mathcal{A}| > 2$, as $|\mathcal{A}|=1$ is trivial and $|\mathcal{A}|=2$ has an inherent symmetry in swapping the labels.
Now the codons graphs as setup as a Hamming graph.
Hamming graphs are known for having automorphisms~\cite{automorph}.
Due to there being a certain number of automorphisms for the Hamming graph for a given $|i|$ and $n$, we quotient~\eqref{eq:config} by the number of symmetries to get the number of unique codes.
As previously noted in Section \ref{subsec:22}, these automorphisms imply that two isomorphic genetic codes should yield the same optimality score.
% \textcolor{blue}{There may be some equation for symmetry factor i couldn't find anything in particular in the paper, some of the terminology was a bit too much for me too understand, but you might be able to}

\subsubsection{Example}\label{subsec:37}
In order to understand the isomorphisms, we will consider an example.
Put $|i|=2$, $n=2$, and $|\mathcal{A}|=3$.
This means $|\mathcal{C}|=4$ giving delta matrices with dimensions $4\times3$.
Using~\eqref{eq:config} we get $\#_{\text{config}}(|\mathcal{C}|=4,|\mathcal{A}|=3)=36$.
We represent this map in the following format:
\begin{equation}
\label{eq:hamming}
\begin{tikzpicture}
\filldraw[black] (0,3) circle (2pt) node[anchor=east]  {01};
\filldraw[black] (3,0)circle (2pt)  node[anchor=west]  {10};
\filldraw[black] (0,0)circle (2pt)  node[anchor=east]  {00};
\filldraw[black] (3,3)circle (2pt)  node[anchor=west]  {11};

\draw[black,thick] (0,3) -- (0,0);
\draw[black,thick] (0,3) -- (3,3);
\draw[black,thick] (3,0) -- (0,0);
\draw[black,thick] (3,0) -- (3,3);

\draw[->,thick] (4,1) -- (8,1);
\filldraw[black] (6,1)node[anchor=south]  {G};

\filldraw[black] (9,2) circle (2pt) node[anchor=east]  {a};
\filldraw[black] (13,2)circle (2pt)  node[anchor=west]  {b};
\filldraw[black] (11,0)circle (2pt)  node[anchor=north]  {c};
\draw[blue,thick] (9,2) -- (13,2);
\draw[blue,thick] (11,0) -- (13,2);
\draw[blue,thick] (9,2) -- (11,0);
\end{tikzpicture}
\end{equation}
In~\eref{eq:hamming}, $G: \{00, 10, 11, 01\}  \mapsto \{c, c, b, a\}$.
We see that the Hamming graph on the left hand side is isomorphic under relabeling~\cite{automorph}. 
In particular, if we relabel $0\longleftrightarrow 1$, the genetic code map would not change.
This configuration has a symmetry factor $18$. 
% \textcolor{blue}{or it should have at least for the 2 surfaces to arise (this is probably wrong but the number of surfaces that appear could just be n sequence length)}.
Taking the quotient of the number of configurations with the symmetry factor suggests that there are only two unique configurations of $\Delta_{c,a}$ for $|\mathcal{C}|=4$ and $|\mathcal{A}|=3$.
Said another way, in this example, there are $4\choose 2$ ways of selecting a pair of codons that are mapped by $G$ to the same amino acid.
Taking into account the repetition, there are $3!$ (the order of $S_3$) ways of mapping the codons to the amino acids.
The product of these terms gives the $36$ configurations.
Taking into account the isomorphisms, we pick out the odd and even elements of the permutation group $S_3$ as our distinguished configurations.

\begin{figure}
\centering
\begin{tabular}{c c}
\includegraphics[width=0.5\linewidth]{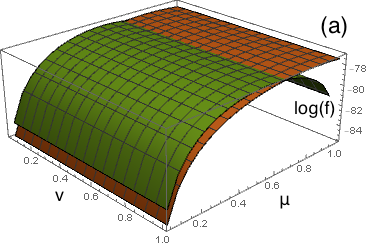}
&
\includegraphics[width=0.5\linewidth]{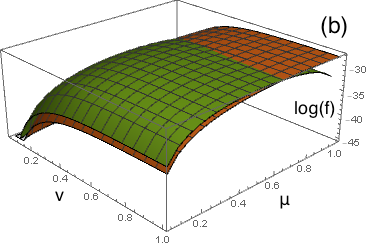}
\\
\includegraphics[width=0.5\linewidth]{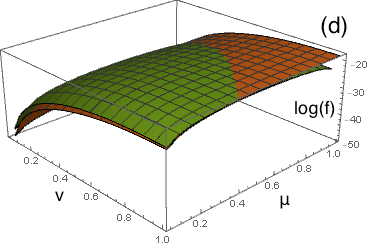}
&
\includegraphics[width=0.5\linewidth]{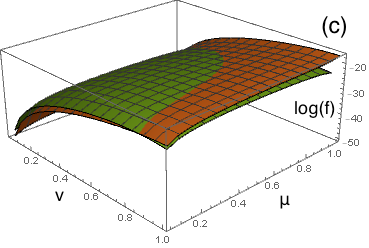}
\end{tabular}

\caption{Four surface plots of $\log f$ in $\mu$--$\nu$ phase space for $\Phi= 0.99$ (a), $0.5$ (b), $0.1$ (c), $0.01$ (d).
We have set $|\mathcal{C}|=4$ and $|\mathcal{A}|=3$. Each colored surface corresponds to a unique genetic code configuration.}
\label{fig:splotsone}
\end{figure}

We will now calculate~\eqref{eq3.2} for all configurations of $\Delta_{c,a}$.
We do this for a given value of $\Phi$ and generate plots in the $\mu$--$\nu$ phase space plane (error space).
In Figure~\ref{fig:splotsone}, we show results for $\Phi= 0.99, 0.5, 0.1, 0.01$.
We expect each unique genetic code configuration to correspond to a unique surface in error space.

\pagebreak
From Figure~\ref{fig:splotsone}, we see that there are two unique surfaces for a given value of $\Phi$.
This is due to there being two unique configurations of $\Delta_{c,a}$.
Within this phase space, there is a curve on which the surfaces interact.
These curves are a critical locus for which the ability of a code to produce a maximum $\log f$ changes.
As $\Phi$ varies, the shape of the surfaces change and critical locus changes.
In the case for $\Phi=0.99$ (Figure~\ref{fig:splotsone} (a)), the critical locus is essentially independent of $\nu$.
The dependence on $\nu$ for the surfaces and the critical locus grow as $\Phi$ decreases.
In order to verify this, we take a polynomial fit to the critical locus for this model.

Taking the ansatz,
\be
\log(f)_\text{fit} = a + b_1\nu + b_2\mu + c_1\nu^2 + c_2\nu\mu + c_3\mu^2 ~,
\ee
for $\Phi=0.99$ the surfaces have polynomials of the form:
\bea
\log(f)_\text{fit} &=& -84.8470+0.0801 \nu-0.0618 \nu^2+22.0655 \mu-15.2344 \mu^2 ~, \label{surf1} \\
\log(f)_\text{fit} &=& -84.0745+0.0522 \nu-0.0404 \nu^2+26.1504 \mu-23.5190 \mu^2 ~, \label{surf2}
\eea
with $R^2 > 0.99995$.
Taking the difference between~\eref{surf1} and~\eref{surf2}, we get the critical locus
\begin{equation}\label{critline}
-0.7725 + 0.0278\nu - 0.0214\nu^2 + 4.0849\mu  +8.2846\mu^2=0 ~.
\end{equation}
As inferred from Figure~\ref{fig:splotsone}, the dependence on $\nu$ is negligible as the coefficients are two or three orders of magnitude smaller than the coefficients for terms involving $\mu$.
For any value of $\Phi$, the coefficient of the $\mu\nu$ cross term ${\cal O}(10^{-10})$.
Thus, at $\Phi\approx 1$,
\begin{equation}\label{eq:assume}
\frac{\partial \log f}{\partial \nu}=0 ~.
\end{equation}
This relation does not necessarily hold for smaller values of $\Phi$ for which we report results in Appendix~\ref{app:e}.
Here, the coefficients of $\nu$ are on a similar magnitude to those for $\mu$. 
This implies that $\Phi$ influences the effects of mistranslations $\nu$ in an inversely proportional manner. 
For the results in the prior sections we use $\Phi=0.99$ for all runs as in~\cite{nigel}. 
This is due to the fact that the effects of mistranslations are more likely to be non-lethal. 
Note that as $\Phi$ and $\mu$ are related through eigenvectors and therefore not linearly related.

Note we also have results for $|i|=4$, $n=1$ and$|\mathcal{A}|=3$ such that we have another case with $|\mathcal{C}|=4$ and $|\mathcal{A}|=3$ but with different automorphisms in Appendix~\ref{app:e}.
For this we find a unique configuration of genetic codes and therefore no critical locus.
We also find that the dependence on $\nu$ increases and $\Phi$ decreases as before.

\subsection{Re-examining universality}\label{subsec:38}
In order to understand the universality of the model in a more formal manner, we examine it in terms of Widom scaling~\cite{widom}.
In particular, we look for homogeneous behavior as a signal of scale invariance on a critical locus.
Consider the logarithm of the fitness function, $\log f$.
As above, $f(\nu,\mu)$ is a function of the rate of mistranslations and the rate of mutations.
Now, homogeneity of $\log f$ demands that
\begin{equation}\label{eq:condition}
\log f(\kappa\nu,\kappa\mu)=\kappa^{\beta}\log f(\nu,\mu) ~,
\end{equation}
where $\kappa\in \mathbb{R}$ is a scale and $\beta$ is the degree of homogeneity.
As $\nu,\mu \in[0,1]$, we require that $\kappa\nu, \kappa\mu \in[0,1]$.
In particular, when $\kappa=1$,
\begin{equation}\label{eq:homog}
\nu \frac{\partial \log f}{\partial \nu}+ \mu\frac{\partial \log f}{\partial \mu}=\beta \log f(\nu,\mu) ~.
\end{equation}
This is the content of Euler's homogeneous function theorem.

However, if we consider the case of $\Phi \approx 1-\epsilon$, where $\epsilon \ll 1$, then contributions from $\nu$ become negligible such that we can apply~\eref{eq:assume}, leaving us to calculate $\frac{\partial \log f}{\partial \mu}$. 
From~\ \eref{eq3.2}, the only part of the $\log f$ that depends on $\mu$ is $U_{c,s}$.
Therefore, we must calculate $\frac{\partial U_{c,s}}{\partial \mu}$. 

Suppose $A$ is a real symmetric matrix with eigenvalues $\lambda_i$ and eigenvectors $\mathbf{v}_i$ such that $\mathbf{v}_i^T \mathbf{v}_i = 1$.
The Perron--Frobenius theorem ensures that the matrix $A$ has a unique real eigenvalue with a magnitude larger than that of any other eigenvalue and a corresponding eigenvector with positive components.
Then
\be
\partial\mathbf{v}_i = (\lambda_i \mathbf{1} - A)^+ (\partial A) \mathbf{v}_i ~,
\ee
where $X^+$ denotes the Moore--Penrose inverse of $X$~\cite{Petersen2008}.
In defining $U_{c,s}$, we have normalized so that $ \sum_{c} U_{c,s} = \mathbf{1}_s$. 
 As $Q_{c,c'}^s$ is a symmetric and real matrix, we therefore only need to rescale $U_{c,s}\rightarrow U_{c,s}'$ such that $\sum_c U_{c,s}' \cdot U_{c,s}'=1$ for any given $s$.
By doing this we can differentiate $U_{c,s}'$:

\begin{equation}\label{eq:order}
\hspace{-0.25cm}
\beta= \frac{\sum_c \sum_s \mu( (\lambda_s^\text{max} \mathbf{1}-Q_{c,c'}^s)^+(\sum_{c''}\frac{\partial M_{c,c''}}{\partial \mu}\delta_{c'',c'}F_{c'',s}) U_{c,s}')'L_{s}}{\sum_c \sum_s U_{c,s}L_{s}} ~,
\end{equation} 
where
\begin{equation}\label{eq:mdiff}
\frac{\partial M_{c,c''}}{\partial \mu}=\left\{
\begin{array}{lcl}
1/(n(|i|-1)) && \text{if  dist}(c,c')=1 ~, \\
-1 && \text{if dist}(c,c')=0 ~, \\
0 && \text{otherwise} ~. \
\end{array}\right.
\end{equation}

We have derived the degree of scaling in the limit $\Phi\to 1$.
This implies that the model is approximately homogeneous in the regime that we have worked in, with degree specified by~\eref{eq:order}.
We see that the order is dependent on the mutations $\mu$, implying that the universality of the code arise due to species having similar mutational errors.
We regard the scaling behavior as a phenomenological observation about the solution near an approximate fixed point of the renormalization group.

We emphasize that when investigating what happens as we tweak the parameters of the model, we calculate the standard deviation to establish which results are significant when we measure the optimality score.
In assessing Widom scaling, we look at the degree of homogeneity, not the degree of optimality; we do this using the fitness score, which is the function actually being maximized in this algorithm, not the optimality score.
The $L_{s}$ terms act as a weighting for each amino acid $s$.
%We can ignore the $L_{s}$ terms, by assuming all amino acids occur with an equal frequency in the proteome.}
We have used a scaling argument in order to justify universality of the algorithm.
The fact that the degree of homogeneity is independent of $\nu$, the rate of mistranslations, in the $\Phi\to 1$ limit, does not imply that the final optimality score is independent of mistranslations.
As seen in Experiment~8 in Section~\ref{subsec:34}, it is not: decreasing $\nu$ increases the optimality score. 
We should emphasise that for smaller values of $\Phi$, \eref{eq:order} will not hold. This is because \eref{eq:assume} begins to break down for values of $\Phi$ outside the condition $\Phi \approx 1$. 
We can amend \eref{eq:order} to consider results around $\Phi=0.9$ by incorporating the addition on an error term through considerations of the first term of \eref{eq:homog}. This approximation does not apply for smaller values of $\Phi$, however, since, as suggested by \citet{sella}, we expect $\Phi$ to be relatively large.
% This result is interesting as we have performed most experiments using $\Phi =0.99$.}

\section{Conclusion and prospects}\label{sec:four}
In this paper, we have argued that with generic initial conditions, there is a late time near universality resulting from the flow of the theory to an attractive solution, \textit{viz.}, the standard genetic code.
The convergence via the attractor mechanism to a near universal solution relies on the mechanism of horizontal gene transfer~\cite{nigel}, which corresponds to setting a parameter $H$ to a non-zero value.
We varied the parameters of the model and found that all variations still display this convergence, except for $H=0$.
This demonstrates the robustness of the model.
We also found that for $0.3< H < 0.7$ we obtain near universal solutions with the greatest degree of optimisation, with $H>0.7$ not being as effective due to some transition from \enquote{mixing} to \enquote{swapping}.
Taking $H$ as a decreasing time-dependent function vastly improves the convergence in comparison to constant values of $H$.
We found that increasing the number of codons, $|\mathcal{C}|$, increases the optimality score $O$.
By limiting the fitness function to a regime that we work in ($\Phi=0.99$), we are able to make approximations that lead to homogeneity in $\log f$, where $f(\nu,\mu)$ is a fitness function depending on the rate of mistranslation and the rate of point mutation.
We derive an expression for the degree of homogeneity, $\beta$.
In the limit $\Phi\to 1$, $\beta$ depends strongly on the mutation rate and negligibly on the mistranslation rate.
We conclude that the rate of point mutations is the crucial factor in driving arbitrary initial conditions to the attractor solution that optimizes fitness of the genetic code.
The point mutation rate is determined by the eigenvalue of the linearized renormalization group transformation around the fixed point for the dynamics.

%In this report we took variations in the setting of parameters for a given algorithm which expresses optimisation to a universal result. We found that increasing the dimensions on the matrices acting on in the algorithm increases the optimality score, while there is an optimal range for the parameter $H$ of $0.3-0.7$  for which we get the most optimal solutions and the fastest rate of convergence. We looked into a criteria for minimising the optimality score $O$ and found that maximising the trace of equation 12 corresponds to minimising the optimality score $O$. We also looked into a real world timescale for this model for which we used to look into how the parameter $H$ might be considered time evolving. The results showed(in figure2) that setting H as some decreasing,time evolving function gives us a very fast rate of convergence especially relative to constant values of $H$. 

% Further investigation of this algorithm would involve testing much larger values for the number of entities $N$.
% We expect the rate of convergence to approach zero as $N$ goes to infinity.
% This is because each iteration will improve $\Delta_{c,a}$ for a single entity.
% Therefore it will take an infinite amount of time steps for us to process all $N$ entities.
% Bifurcations around $H=0$ are likely, as the results suggest convergence towards a universal solution only for $H\neq 0$.
% Also further investigation into variations on the parameters $|\mathcal{A}|$, $a_d$, and $ L_{s}$ would be beneficial.

Improvements to make the algorithm more accurate for biology would involve exploring how to incorporate stop codons which do not code for amino acids into the model as something more than a dummy amino acid.
We should also consider that mutations and horizontal gene transfer do not occur at the same rate as suggested by  both occurring at each iteration.
Some work to estimate a timescale for this model possibly by considering rate of error as the sum of all errors ($\nu+\mu$) and relating this to  the measured rate of error in, for example, a kinetic proofreading model~\cite{alon2019introduction}.
We also suggest that additional factors and steps should be incorporated in order to guarantee a universal solution is converged to every run.

Variations on the dimensions, $|\mathcal{C}|$ and $|\mathcal{A}|$, which count the number of codons and the number of amino acids, respectively, display convergence to a universal result.
This could have applications to synthetic biology  where codes with up to $8$ bases have been created~\cite{Hoshika884}. 
These codes should also converge to a universal genetic code given enough time.
An open question is to determine what sets the initial conditions.
Why did life on Earth evolve to make use of four base pairs in DNA, three base pairs per codon, and $20$ amino acids?

We have focused on a single basin of attraction, whereas there could be others.
The basin of attraction may be determined by biochemistry inputs.
We can imagine, for example, a different basin of attraction in which the solvent is ammonia, methane, or hydrogen fluoride instead of water.
We can also imagine biochemistry organized around silicon instead of carbon.
The molecular realization of the genetic code would be different based on these other inputs, but we expect that the same principles apply, and these other hypothetical genetic codes would also evolve to a universal solution based on the principle of horizontal gene transfer.

Broadly speaking, we have argued that the concept of universality from statistical physics applies to biological systems like the genetic code.
The thermodynamic limit arises from a large $N$ number of degrees of freedom in the entities studied.
The dynamical system is driven to an attractor solution as a result of interactions, in this case horizontal gene transfer.
We have considered a mechanism for horizontal gene transfer and by tweaking its parameters identified which ones are the most important.
The existence of approximate homogeneity offers evidence for universality.
We would like to interrogate how general this setup is and whether it is useful for studying other complex systems.

Indeed, like thermodynamics and evolution itself, we wish to consider horizontal gene transfer as an organizing principle in Nature.
Different solutions to a theory or different possible initial conditions can exchange information with each other through complex processes.
Dynamics can then flow the system to a late time attractor solution that is independent of specific parameters of the model.
As a proving ground for this hypothesis, we can consider the vacuum selection problem in quantum gravity.
String theory, a promising candidate framework for marrying gravitation with quantum theory, generically predicts a landscape of vacua, one of which is our Universe with the Standard Models of particle physics and cosmology as phenomenological features that explain dynamics at small and large scales.
(See, for example,~\cite{Denef:2006ad, jkm, Yang} for related reviews.)
These vacua inevitably arise as a consequence of a simple observation --- we live in four spacetime dimensions whereas the consistency of the theory demands ten, and there is no unique way to reduce the number of dimensions.
Because they are unobserved, the extra dimensions predicted by string theory comprise a compact geometry with special properties.
The moduli space of string compactifications is believed to be connected.
We can calculate the degree of fine tuning necessary to support certain cosmological structures and the astrophysical and chemical preconditions necessary for life~\cite{Adams:2019kby}.
Rather than making an explicitly anthropic argument~\cite{Weinberg:1988cp, Polchinski:2006gy}, we can model a dynamics for vacuum selection which incorporates a mechanism analogous to horizontal gene transfer to lead to universal and optimal structures as an attractive fixed point.
Thus, instead of arguing that low energy observables such as the cosmological constant are distributed randomly across the landscape, horizontal gene transfer, by driving the system to the attractor value, may obviate aspects of the measure problem.
Developing and testing this hypothesis within the string theory framework is work in progress.

%Indeed, like thermodynamics and evolution itself, we regard horizontal gene transfer as an organizing principle in nature.
%String theory predicts a landscape of vacua, one of which is our Universe with the Standard Models of particle physics and cosmology as phenomenological features that explain dynamics at small and large scales.
%(See, for example,~\cite{Denef:2006ad, jkm, Yang} for related reviews.)
%We can calculate the degree of fine tuning necessary to support certain cosmological structures and the astrophysical and chemical preconditions necessary for life~\cite{Adams:2019kby}.
%Rather than making an explicitly anthropic argument~\cite{Weinberg:1988cp, Polchinski:2006gy}, we can test whether  an interaction modeled on horizontal gene transfer can lead to universal and optimal structures as an attractive fixed point.
%This is work in progress.

\section*{Acknowledgements}
We are grateful to David Ardell, Nigel Goldenfeld, Sujay Nair, and Kalin Vetsigian for feedback and insightful discussions.
YHH is indebted to the Science and Technology Facilities Council, UK, for grant ST/J00037X/1.
VJ thanks the South African Research Chairs Initiative of the Department of Science and Technology and the National Research Foundation for support.
DM thanks the Julian Schwinger Foundation and the US Department of Energy (DE-SC0020262) for support.

\appendix

%\vj{Fix internal captions in Figures~6, 7, 8.
%These shouldn't be hard coded anyway.}
\newpage
\section{Different initial delta matrices $\Delta_{c,a}$}\label{app:a}
\vspace{-0.1cm}
Initializing with different delta matrices, the optimality score converges.
%\begin{center}
\vspace{-0.1cm}
\begin{figure}[H]
\centering
\begin{tabular}{c c}
\includegraphics[width=0.5\linewidth]{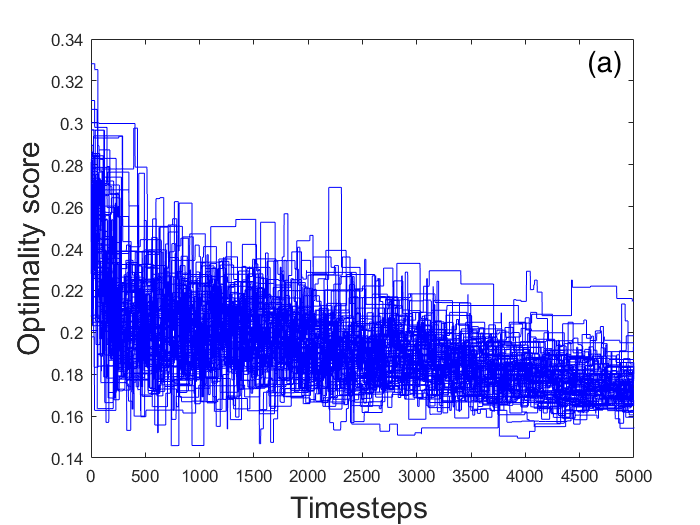}
&
\includegraphics[width=0.47\linewidth]{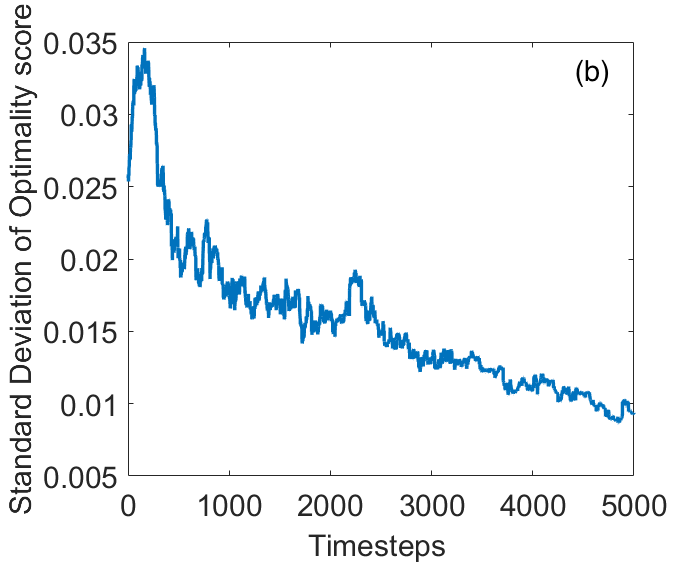}
\end{tabular}
\caption{Graph showing evolution of optimality score when all entities have a different initial $\Delta_{c,a}$  rather than the same initial $\Delta_{c,a}$. 
Initial parameters are: $|i|=3$, $n=3$, $|A|=9$, $N=80$, $K=1$, $H=0.4$, $\nu=0.01$, $\mu=10^{-4}$, and $ \Phi=0.99$}
\end{figure}
\pagebreak

\section{Varying innovation pool structure}\label{app:b}
\vspace{-0.1cm}
As discussed in Section~\ref{subsec:33}, we plot what happens as we vary $N$, the number of entities under consideration, and $K$, the number of donors.
\vspace{-0.1cm}

\begin{figure}[H]
\centering
\begin{tabular}{c c}
\includegraphics[width=0.5\linewidth]{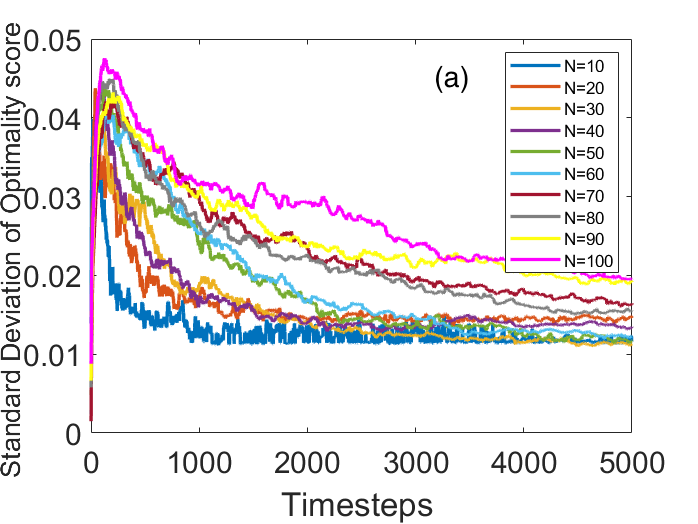}
&
\includegraphics[width=0.5\linewidth]{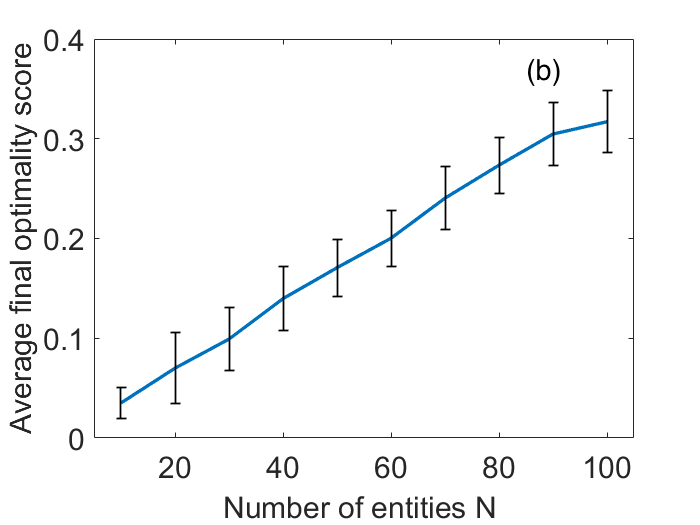}
\end{tabular}

\caption{Figure \ref{fig:entity} (a) shows the average time evolution of the standard deviation of the optimality score for a given $N$ over ten runs. 
Figure \ref{fig:entity} (b) shows $N$ (number of entities) against the average final optimality score, averaged over ten runs.
The initial parameters are the same for all runs: $|i|=3$, $n=3$, $|\mathcal{A}|=9$, $K=1$ , $H=0.4$, $\nu=0.01$, $\mu=10^{-4}$, and $ \Phi=0.99$.
The error bars show the average one standard deviation spread of final optimality scores over ten runs (to measure the rate of convergence).}
\label{fig:entity}
\end{figure}

\begin{figure}[H]
\vspace{1cm}
\centering
\begin{tabular}{c c}
\includegraphics[width=0.47\linewidth]{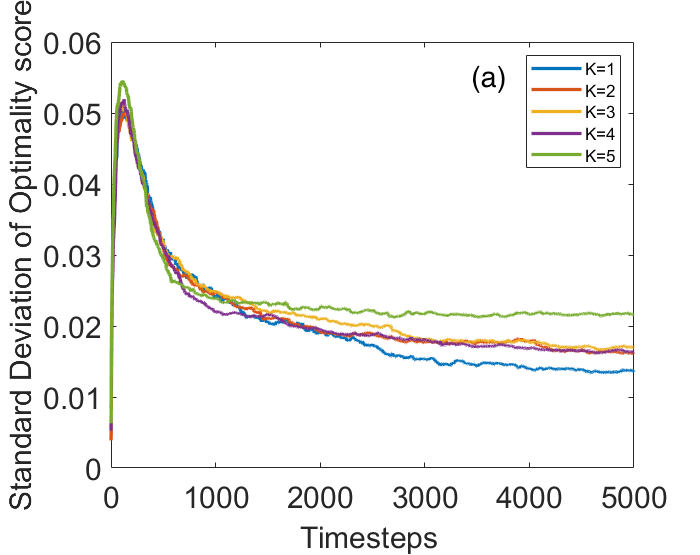}
&
\includegraphics[width=0.5\linewidth]{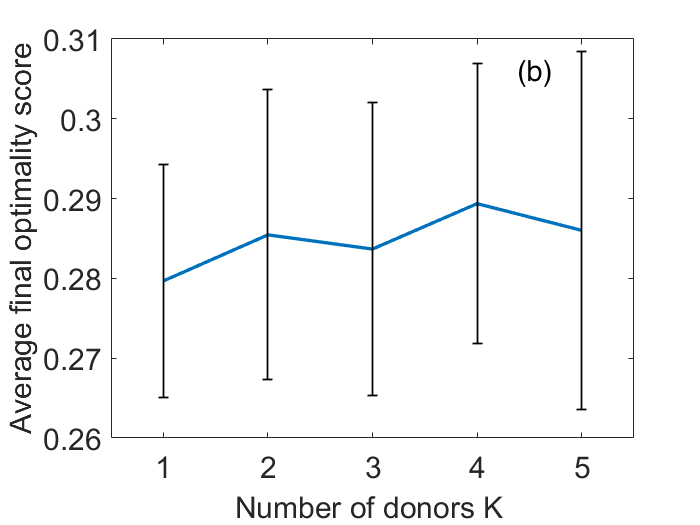}
\end{tabular}

\caption{Figure \ref{fig:donor} (a) shows the average time evolution of the standard deviation of the optimality score for a given $K$ over ten runs. 
Figure \ref{fig:donor} (b) shows $K$ (number of donors) against the average final optimality score, averaged over ten runs.
The initial parameters are the same for all runs:  $|i|=3$, $n=3$, $|\mathcal{A}|=9$, $N=80$, $H=0.4$, $\nu=0.01$, $\mu=10^{-4}$, and $ \Phi=0.99$.
The error bars show the average one standard deviation spread of final optimality scores over ten runs (to measure the rate of convergence).}
\label{fig:donor}
\end{figure}

\newpage

\section{Time evolution of H}\label{app:evo}
\vspace{-0.1cm}
%As discussed in experiment 7 in Section~\ref{subsec:33}, We look at the results for a time evolving $H$.
\vspace{-0.1cm}
\begin{figure}[h!!!]
\centering
\begin{tabular}{c c}
\includegraphics[width=0.5\linewidth]{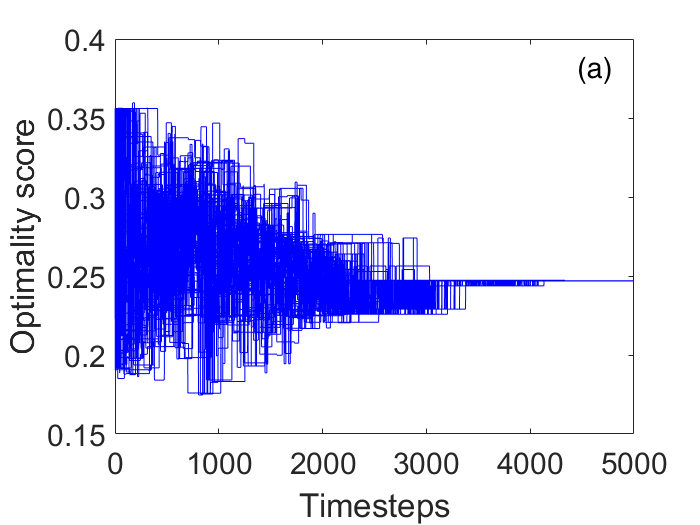}
&
\includegraphics[width=0.5\linewidth]{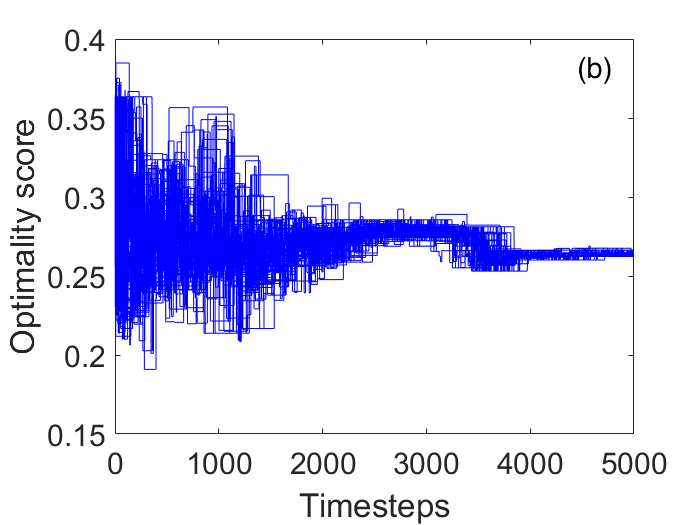}
\\
\includegraphics[width=0.5\linewidth]{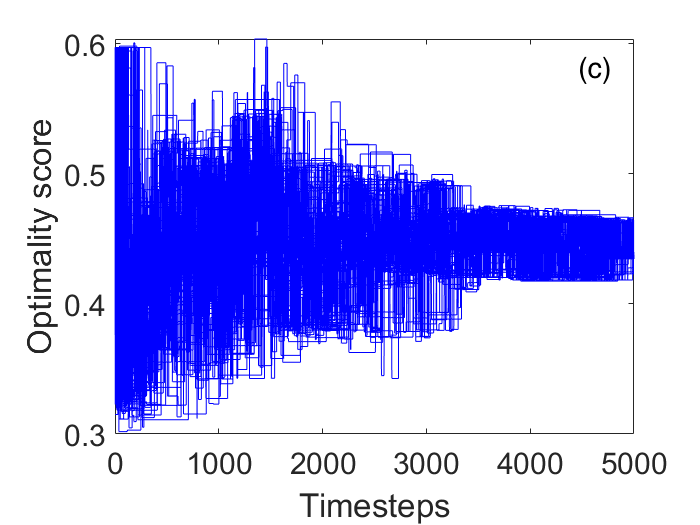}
&
 \includegraphics[width=0.5\linewidth]{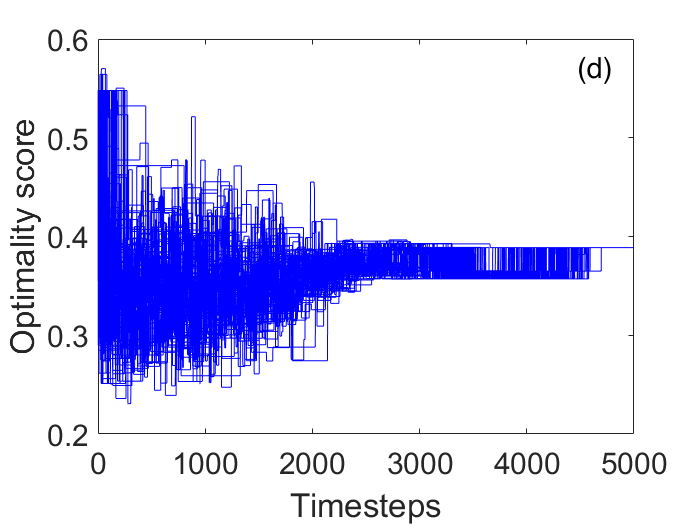}
 \\
\includegraphics[width=0.5\linewidth]{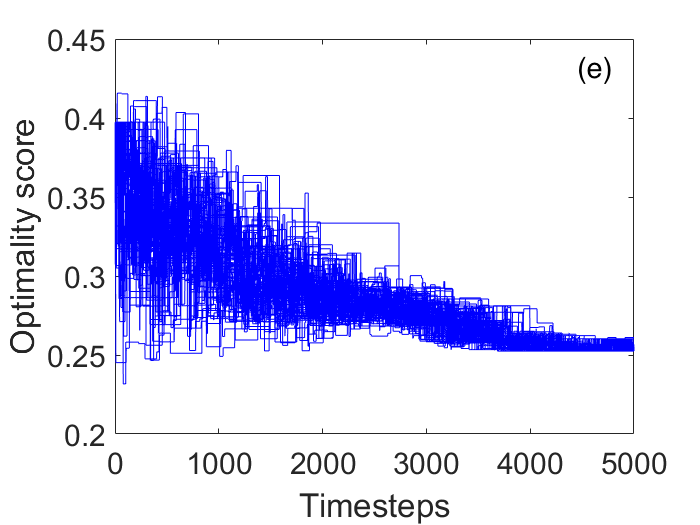}
&
 \includegraphics[width=0.5\linewidth]{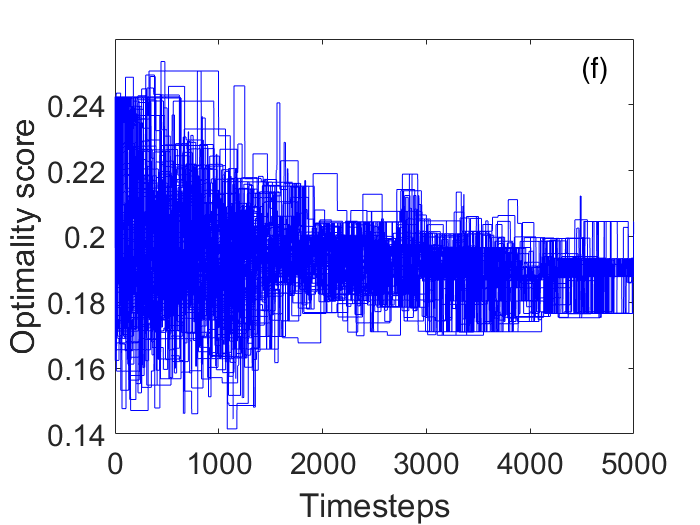}
\end{tabular}
\caption{Evolution of optimality score with a time evolving parameter $H$.
Six runs showing the evolution of optimality score for a time evolving $H$ (according to \eqref{eq:timeh}).
We use: $|i|=3$, $n=3$, $|A|=9$, $N=80$, $K=1$, $\nu=0.01$, $\mu=10^{-4}$, $ \Phi=0.99$, $k=10^{-4}$, and $H_0=1$.}
\label{fig:timeHapp}
\end{figure}

\newpage

\section{Varying noise and fitness parameters}\label{app:c}
\vspace{-0.1cm}
As discussed in Section~\ref{subsec:34}, we plot variations of $\nu$ (the mistranslation rate), $\mu$ (the mutation rate), and $\Phi$ (the scale for abstract physicochemical distance).% between amino acids}).
\vspace{-0.1cm}

\begin{figure}[H]
\centering
\begin{tabular}{c c}
\includegraphics[width=0.495\linewidth]{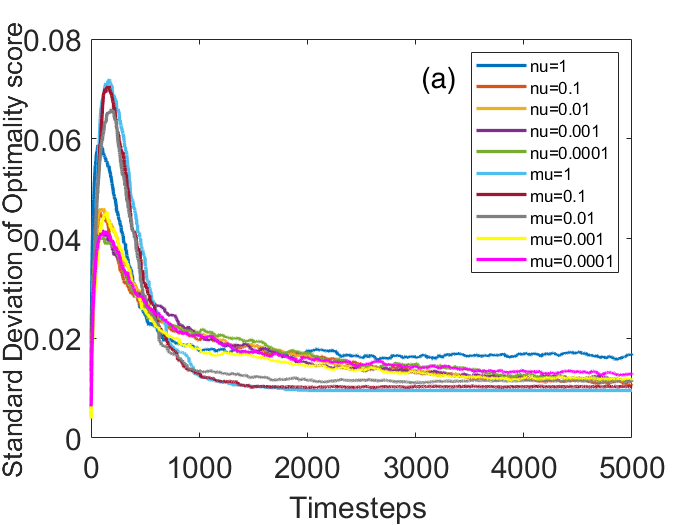}
&
\includegraphics[width=0.495\linewidth]{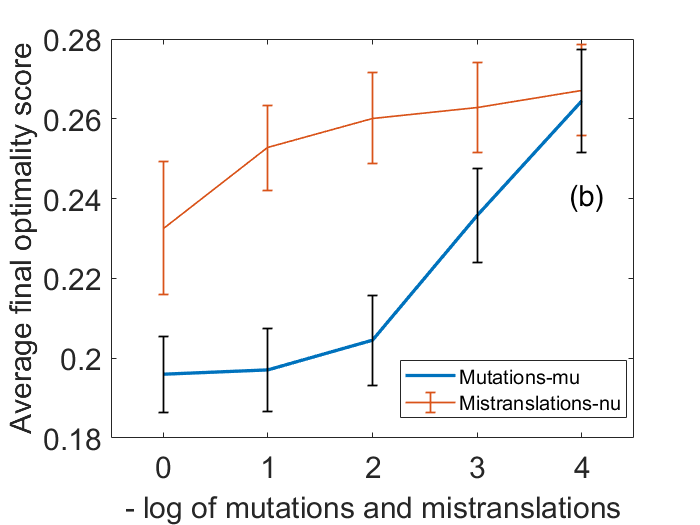}
\end{tabular}
\caption{Figure \ref{fig:error} (a) shows the average time evolution of the standard deviation of the optimality score for a given $\nu$ and $\mu$ over ten runs. 
Figure \ref{fig:error} (b) shows $\nu$ and $\mu$ against the average final optimality score, averaged over ten runs.
The initial parameters are the same for all runs:  $|i|=3$, $n=3$, $|\mathcal{A}|=9$, $N=80$, $K=1$, $H=0.4$, and $ \Phi=0.99$.
When varying $\nu$, we put $\mu=10^{-4}$.
When varying $\mu$, we put $\nu=0.01$. 
The error bars show the average one standard deviation spread of final optimality scores over three runs (to measure the rate of convergence).}
\label{fig:error}
\end{figure}

\vspace{-0.2cm}
\begin{figure}[H]
\centering
\begin{tabular}{c c}
\includegraphics[width=0.47\linewidth]{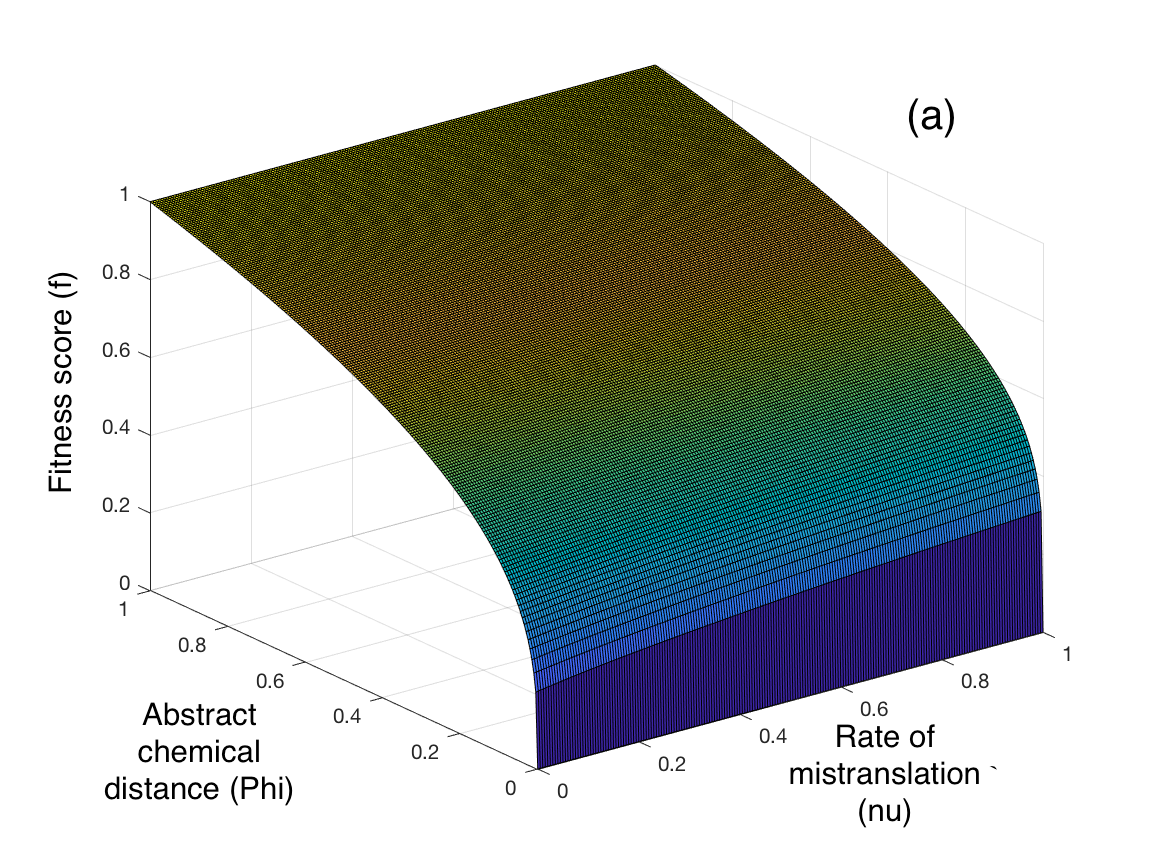}
&
\includegraphics[width=0.5\linewidth]{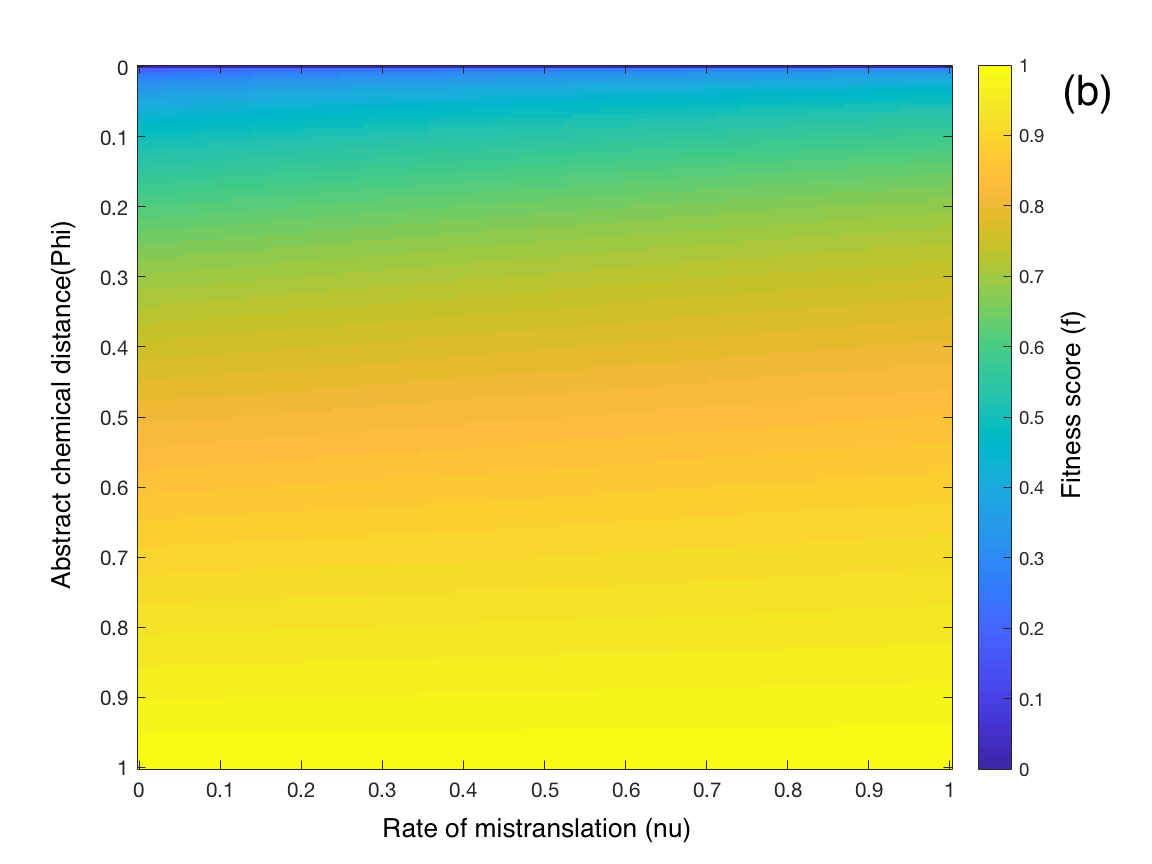}
\\
\end{tabular}
\caption{Figure \ref{fig:phi} (a) displays a surface plot of the fitness function $f$ in terms of $\Phi$ and $\nu$.
Figure \ref{fig:phi} (b) is simply a heatmap of the first plot.
We use: $|i|=3$, $n=3$, $|\mathcal{A}|=9$, $N=80$, $K=1$, $H=0.4$, and $\mu=10^{-4}$.
Other parameters are randomly generated.}
\label{fig:phi}
\end{figure}
\vspace{-0.1cm}
\section{Defining universality}\label{app:e}
We show the best fits for different values of $\Phi$ corresponding to the surfaces for $\log f$ as a function of $\nu$ and $\mu$ in Figure~\ref{fig:splotsone}.
\begin{figure}[H]
\centering
\begin{tabular}{c c}
\includegraphics[width=0.5\linewidth]{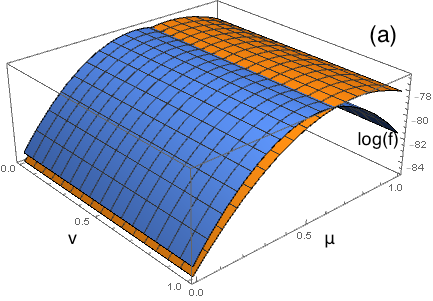}
&
\includegraphics[width=0.5\linewidth]{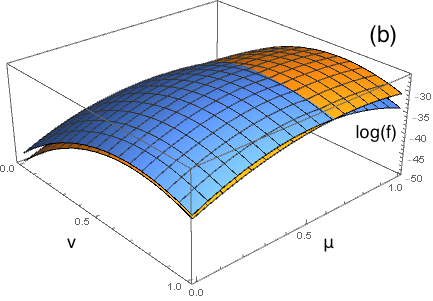}
\\
\includegraphics[width=0.5\linewidth]{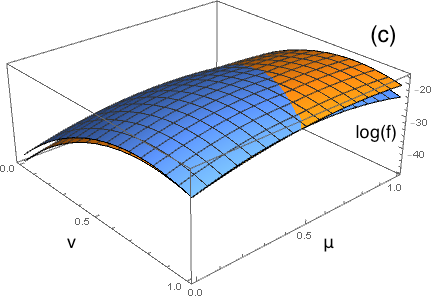}
&
\includegraphics[width=0.5\linewidth]{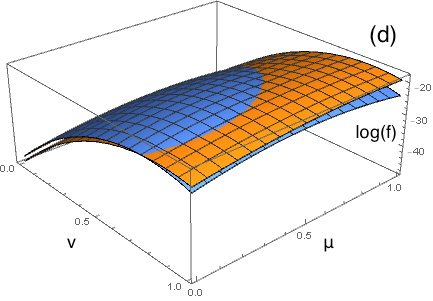}
\end{tabular}
\caption{Four surface plots of polynomial fit of $\log f$ in $\mu$--$\nu$ phase space for $\Phi= 0.99$ (a), $0.5$ (b), $0.1$ (c), $0.01$ (d).
This is to be compared with Figure~\ref{fig:splotsone}.}
\label{fig:fits}
\end{figure}

In analogy to~\eref{surf1} and~\eref{surf2} for $\Phi=0.,99$, we provide polynomial best fits for $|i|=2$, $n=2$ and $|\mathcal{A}|=3$ for $\Phi = 0.5, 0.1, 0.01$.

\begin{itemize}
\item \textbf{$\Phi=0.5$:}

\begin{equation}
\log(f)_\text{fit}=-49.6633+44.1899 \nu-31.6042 \nu^2+21.4761 \mu-14.8319 \mu^2\ , \qquad R^2=0.99761 ~.
\end{equation}

\begin{equation}
\log(f)_\text{fit}=  -47.5384+40.8086 \nu-29.5881 \nu^2+25.3872 \mu-22.8529 \mu^2\ , \qquad R^2=0.997706 ~.
\end{equation}

\item \textbf{$\Phi=0.1$:}

\begin{equation}
\log(f)_\text{fit}=  -47.7527+61.5566 \nu-39.4707 \nu^2+16.8343 \mu-11.4213 \mu^2\ , \qquad R^2= 0.99705 ~,
\end{equation}

\begin{equation}
\log(f)_\text{fit}= -45.5857+59.7551 \nu-39.8311 \nu^2+19.962 \mu-18.1142 \mu^2\ , \qquad R^2=0.996994 ~.
\end{equation}

\item \textbf{$\Phi=0.01$:}

\begin{equation}
\log(f)_\text{fit}= -48.6657+62.1467 \nu-36.9472 \nu^2+11.0255 \mu-6.69862 \mu^2\ , \qquad R^2=0.997404 ~,
\end{equation}

\begin{equation}
\log(f)_\text{fit}= -46.9935+63.104 \nu-41.1589 \nu^2+14.3666 \mu-13.2086 \mu^2\ , \qquad R^2= 0.997725 ~.
\end{equation}

\end{itemize}

\begin{figure}[H]
\centering
\begin{tabular}{c c}
\includegraphics[width=0.5\linewidth]{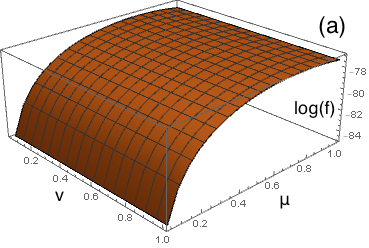}
&
\includegraphics[width=0.5\linewidth]{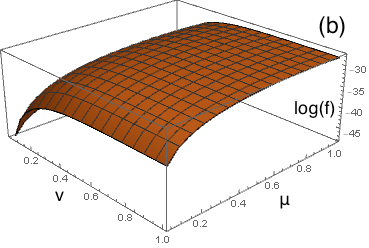}
\\
\includegraphics[width=0.5\linewidth]{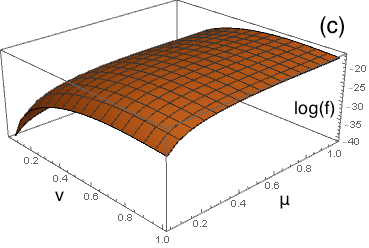}
&
\includegraphics[width=0.5\linewidth]{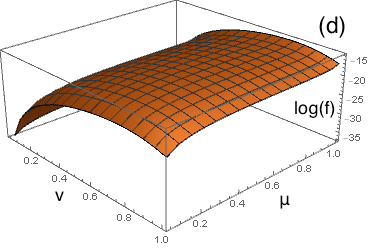}
\end{tabular}

\caption{Four surface plots of polynomial fit of $\log f$ in $\mu$--$\nu$ phase space for $\Phi= 0.99$ (a), $0.5$ (b), $0.1$ (c), $0.01$ (d).}
\label{fig:extratest}
\end{figure}

The values $|i|=4$, $n=1$, $|\mathcal{C}|=4$, $|\mathcal{A}|=3$ is represented by the map:
\begin{equation}
\begin{tikzpicture}
\filldraw[black] (0,3) circle (2pt) node[anchor=east]  {4};
\filldraw[black] (3,0)circle (2pt)  node[anchor=west]  {2};
\filldraw[black] (0,0)circle (2pt)  node[anchor=east]  {1};
\filldraw[black] (3,3)circle (2pt)  node[anchor=west]  {3};

\draw[black,thick] (0,3) -- (0,0);
\draw[black,thick] (0,3) -- (3,3);
\draw[black,thick] (3,0) -- (0,0);
\draw[black,thick] (3,0) -- (3,3);
\draw[black,thick] (0,0) -- (3,3);
\draw[black,thick] (3,0) -- (0,3);

\draw[->,thick] (4,1) -- (8,1);
\filldraw[black] (6,1)node[anchor=south]  {G};

\filldraw[black] (9,2) circle (2pt) node[anchor=east]  {a};
\filldraw[black] (13,2)circle (2pt)  node[anchor=west]  {b};
\filldraw[black] (11,0)circle (2pt)  node[anchor=north]  {c};
\draw[blue,thick] (9,2) -- (13,2);
\draw[blue,thick] (11,0) -- (13,2);
\draw[blue,thick] (9,2) -- (11,0);

\end{tikzpicture}
\end{equation}
These give the surface plots of polynomial fit of $\log f$ over the $\mu$--$\nu$ phase space shown in Figure~\ref{fig:extratest}.

%{\LARGE \begin{center} {\color{blue}{\textbf{--- END OF VJ EDIT ---}}} \end{center}}

\newpage
\bibliographystyle{unsrtnat}%plainnat}
\bibliography{references}

\end{document}